\documentclass[superscriptaddress,11pt]{revtex4-2}
\usepackage{graphicx}
\usepackage{subfig,color}
\usepackage{natbib}
\usepackage{pdfpages}
\setcitestyle{super}
\usepackage{url}
\usepackage{pifont}
\usepackage{amsmath}
\usepackage{float}
\usepackage{placeins}
\usepackage{changepage}
\usepackage{xcolor}
\usepackage{multibib}
\usepackage[symbol]{footmisc}
\newcites{supp}{Supplementary}

\usepackage{xr-hyper}  % Use xr-hyper for compatibility with hyperref
\usepackage[hidelinks]{hyperref}

\makeatletter
\AtBeginDocument{\let\LS@rot\@undefined}
\makeatother

\usepackage[counterclockwise, figuresright]{rotating}

\usepackage{adjustbox}
\usepackage[normalem]{ulem}

\def\revtxt{black}
\def\revttxt{black}
\def\be{\begin{equation}}
\def\ee{\end{equation}}

\def\begineqn{\begin{equation*}}
\def\endeqn{\end{equation*}}

\def\beginar{\begin{eqnarray}}
\def\endar{\end{eqnarray}}
\def\beginarn{\begin{eqnarray*}}
\def\endarn{\end{eqnarray*}}

\def\lb{\left ( }
\def\rb{\right ) }
\def\ls{\left[ \vphantom] }
\def\rs{\right] }

\def\ep{\epsilon}

\def\Rat{\widetilde{Ra}}

\def\ub{\mathbf{u}}

\def\tt{\widetilde{t}}
\def\ot{\widetilde{\omega}}

\def\Gtz{{ {\Gamma_z} }}
\def\Gstd{{ {\Gamma}^{\prime}_z }}

\def\Gm{{\Gamma}}
\def\ricb{{ {r_{icb}}}}
\def\rcmb{{ {r_{cmb}}}}
\def\ticb{{ {T_{icb}}}}
\def\tcmb{{ {T_{cmb}}}}

\def\de{\delta_{\text{E}}}

\begin{document}
\renewcommand{\figurename}{Fig.}
\renewcommand{\tablename}{Table}

%\textbf{\huge{Cover Letter}}
%
%The submitted article presents the best evidence to date that topographic coupling between the outer core and the mantle of the Earth can explain the sub-decadal variation in the Earth’s Length of Day. Until now, evidence for topographic coupling has been limited to untested, theoretical scaling arguments and a single set of simulations at mild, non-geophysical conditions. The data reported in this article represents a large set of state-of-the art simulations which demonstrate that topographic coupling satisfies the necessary conditions for the sub-decadal Length of Day variations. Furthermore, the strategies used in this study to model topography provide a new paradigm for numerical studies of convection in geo- and astro-physical bodies with non-trivial topographic structures. 
%
%The reported results were presented in a poster format at the 2024 Study of the Earth’s Deep Interior (SEDI) conference in Great Barrington, Massachusetts, USA. We report no other dissemination.
%We report no conflicts of interest.
%\thispagestyle{empty}
%\newpage

\clearpage
\pagenumbering{arabic}
%\title{The influence of boundary topography on outer core dynamics}
%\title{Topographic torques as a possible cause for decadal variations in the length of day}
%\title{Quantifying topographic torques at the core mantle boundary with direct numerical simulations}
%\title{Large topographic torques at the core mantle boundary and Earth's length of day: convection in a rotating dimpled sphere}
%\title{Convection in a rotating deformed sphere: large topographic torques at Earth's core mantle boundary}
%\title{Large topographic torques at Earth's core mantle boundary and variations in the length of day}
\title{Turbulence in Earth's core generates large topographic torques on the mantle}

\author{Tobias G. Oliver\footnote{tobias.oliver@colorado.edu}}
\affiliation{Department of Physics, University of Colorado Boulder, Boulder, CO, USA}

\author{Eric G. Blackman}
\affiliation{Department of Physics and Astronomy, University of Rochester, Rochester, NY, USA}

\author{John A. Tarduno}
\affiliation{Department of Earth and Environmental Sciences, University of Rochester, Rochester, NY, USA}

\author{Michael A. Calkins}
\affiliation{Department of Physics, University of Colorado Boulder, Boulder, CO, USA}
\begin{abstract}
	Seismic and geodynamic studies indicate that the boundary between the Earth's liquid outer core and solid mantle is not spherical, but is likely characterized by topography in the form of inverted mountains and valleys that have typical amplitudes of several kilometers.  
	One of the dynamical consequences of these deformations is that turbulent flow in the core can exert pressure torques on the mantle, thereby resulting in a transfer of angular momentum between the outer core and the mantle. Understanding this transfer of angular momentum is important for explaining variations in the Earth's rotation rate, or length of day.
	Whether kilometer-sized topography can explain observed variations in length of day is a longstanding question in geophysics. 
	\textcolor{\revtxt}{Here we use a suite of state-of-the-art numerical simulations of hydrodynamic convection in a rotating spherical shell with boundary topography to show that topographic torques exhibit a linear dependence on topographic amplitude and approach a quadratic dependence on flow speeds. This observation is explained with the asymptotic theory of rapidly rotating convection.
%Hence, topographic torques satisfy the primary requirements needed to explain observed length of day variations on decadal time scales.
These results imply that topographic torques are of sufficient magnitude to explain length of day variations.}

\end{abstract}
%\resetlinenumber
\maketitle

\section*{Introduction}
\label{S:intro}

The Earth's rotation rate, or length of day (LOD), is known to vary over a broad spectrum of frequencies, ranging from the slow, secular increase caused by tidal dissipation, to inter-annual variations resulting from the interaction between atmospheric flows and the solid crust \cite[e.g.][]{jW88,kL05}. \textcolor{\revtxt}{Large amplitude fluctuations are present in measurements of the LOD, ranging from timescales of several decades ($\sim 60$) \citep{rH98} to sub-decadal ($\sim 6$ yr) \citep{rH15}. These fluctuations} are widely believed to be the result of exchanges of angular momentum between the liquid core and the solid mantle \citep{dJ88,aJ93,rH15}, however the physical \textcolor{\revtxt}{mechanisms responsible for this exchange are not fully understood \citep[e.g.][]{pR12}. The required axial torques} may arise from a variety of sources including pressure gradients, electromagnetic and viscous stresses at the core mantle boundary (CMB), as well as gravitational torques between the mantle and solid inner core \citep{bB96}. 
\textcolor{\revtxt}{Previous scaling arguments \citep[e.g.][]{pR12} suggest that viscous drag on the CMB is too small to produce the required torque. 
}
The electromagnetic torques can explain LOD  if the electrical conductivity of the lower mantle is sufficiently large \citep[e.g.][]{rH98}. 
	Gravitational coupling between the inner core and mantle is also capable of producing the target torque of $10^{18} \text{N}\cdot \text{m} $ \citep{bB96,bB09}.
Pressure, or topographic, coupling occurs wherever the CMB deviates from spherical symmetry. 
%%REMOVED BY TO 101024
%In the context of changes in LOD, a rotational asymmetry about the rotation axis of the Earth is required -- variations of pressure over the surface of the topography can then yield non-zero torques on the mantle \citep{rH69}. 
Topographic coupling between the core and mantle is one of the least understood mechanisms, due in part to the complex techniques required to model it. \textcolor{\revttxt}{To the author's knowledge}, only a single study has used three-dimensional simulations, and this work used an approximate set of boundary conditions over a limited range of parameters \citep{wK01}. 
Towards this end, the present work examines the role of topographic coupling over a broad set of numerical experiments of rotating convection in a spherical shell in which topography is superposed onto the outer boundary. The primary goals are to test the scaling behavior of topographic torques with varying input parameters, including the rotation rate, forcing strength, and topographic amplitude, and extrapolate the findings to the conditions of the outer core.

Seismological studies report CMB topography with peak to trough amplitudes near $5$km \citep{sT10,kK03}.
There is little overlap in the spatial structure of seismically-derived CMB topography, indicating that it is poorly constrained \citep[e.g.][]{pK21}. 
Nevertheless, it is likely that CMB topography is characterized by a broad spectrum of amplitudes and wavelengths, and may therefore have a structure that is similar to the Earth's surface \citep[e.g.][]{aE18,mP23}, albeit with a smaller amplitude. 
%%REMOVED BY TO 101024
%Indeed, just as the mantle and plate tectonics drive the formation of topographic features at the surface, motions in the lower mantle likely shape the CMB. 
The presence of seismically slow material at the base of the mantle (e.g.~the so-called Large Low Shear Velocity Provinces, or LLSVPs) below southern Africa and the southern Pacific \citep[e.g.][]{sC16} likely contribute to deformations at the CMB which may influence flow in the outer core\citep{jT15}. 
%%REMOVED BY TO 101024
%Indeed, geodynamic simulations of mantle convection show that such structures readily lead to topography at the CMB, though the shape and amplitude of the topography is dependent on the particular thermal and compositional nature of the LLSVPs \citep{tL07,tL10}. 

When deviations in spherical symmetry are present on the CMB, non-radial pressure forces 
will be exerted on the CMB by the outer core. 
Such forces can then give rise to torques on the mantle that, in turn, cause changes in its angular momentum. \textcolor{\revtxt}{This topographic torque is defined by
\begin{equation}
	\boldsymbol{\Gm} = \oint  \mathbf{r}\times  p\; d\mathbf{a},
	\label{e:top_torque}
\end{equation}
where $\mathbf{r}$ is the position vector pointing from the center of the planet to the surface of the CMB, $p$ is pressure and $d\mathbf{a}$ is a surface area element of the CMB with the normal chosen to point away from the core. In this study we allow for deviations from spherical symmetry and denote the radial position of the CMB as $r = \rcmb -  \epsilon \, h(\theta, \phi)$, where $\rcmb = 20/13$ is the non-dimensional radius of the CMB in the absence of topography, $\epsilon$ is the topographic amplitude and $h(\theta, \phi)$ defines the topographic shape (see equation \ref{e:h}). In our non-dimensionalization, $\epsilon$ is the ratio of topography depth (in km) to the core depth (2260 km) and is therefore expected to be $O(10^{-3})$ at the CMB. 
If we assume that topographic amplitudes are small relative to the depth of the core and some characteristic topographic width, an approximate expression for the axial component (i.e.~pointing in the direction of Earth's rotation axis) of the topographic torque is given by
\be
\Gtz \approx \epsilon \oint h\lb \theta,\phi\rb \partial_\phi p \; d a, 
\label{E:torque}
\ee
where $\partial_{\phi} p$ is the longitudinal pressure gradient within the core.}
If the pressure variations are induced by the topography, then \textcolor{\revtxt}{there is an order $O(\ep)$ perturbation to the pressure gradient and} $\Gtz = O\lb \epsilon^{2}\rb $\textcolor{\revtxt}{\citep{aA77b,wK93,sG18,dJ20,mP23}}. 
Alternatively, if the dominant contribution to the pressure gradient is not dependent on the topographic amplitude, the magnitude of the axial torque scales like $\Gtz = O(\epsilon)$.
Given that $\ep \sim 10^{-3}$ at the CMB, these two scalings can yield vastly different estimates for the topographic torque. 
\textcolor{\revtxt}{
	We ensure that the topographic amplitude is less than the topographic width in all of our simulations so that these two mechanisms can be distinguished, although we emphasize that the exact relation (equation \ref{e:top_torque}) is used to calculate the torques. Details on the topographic parameters are reported in the Methods (Table \ref{t:bump_geo}).
}
\textcolor{\revttxt}{We are unaware of any previous studies that} have conclusively demonstrated the existence of either scaling in a fully self-consistent convection-driven flow.

%Predictions for $\Gamma_{z},$ the topographic torque about the Earth's rotation axis, can be expressed as \textcolor{\revtxt}{
%$\Gtz \approx \epsilon \oint h\lb \theta,\phi\rb \partial_\phi p \; d a$\cite{pR12,rH69}, 
%where $\epsilon$ is the topographic amplitude non-dimensionalized by the core thickness 2260 km and $\partial_{\phi} p$ is the longitudinal pressure gradient.
%}
%%in terms of $\epsilon$ and $\partial_{\phi} p,$ the non-dimensional topographic amplitude and azimuthal (longitudinal) pressure gradient 
%%\cite{pR12,rH69}. 
%If the pressure variations are induced by the topography, then $\Gtz = O\lb \epsilon^{2}\rb $\citep{aA77b,wK93,sG18,dJ20}. 
%Alternatively, if the dominant contribution to the pressure gradient is not dependent on the topographic amplitude, the magnitude of the axial torque scales like $\Gtz = O(\epsilon)$.
%Given that $\ep \sim 10^{-3}$ at the CMB, these two scalings can yield vastly different estimates for the topographic torque. 
% To date, no previous studies have conclusively demonstrated the existence of either scaling in a fully self-consistent convection-driven flow. Here we confirm that the linear scaling is appropriate in the core, indicating that small scale convective turbulence is capable of generating topographic torques of sufficient magnitude to explain sub-decadal LOD variations. 

 We address the problem of core mantle topographic coupling by employing direct numerical simulations of \textcolor{\revtxt}{hydrodynamic} convection in a rotating spherical shell in which a single topographic bump with Gaussian profile protrudes into the core; a schematic of the domain is shown in \textcolor{\revttxt}{Supplementary} Fig. \ref{f:topo}. This choice of topography is made both for simplicity, and to ensure a rich interaction between core flows and the mantle given that a Gaussian profile is characterized by a broad spectrum of wavelengths. 
 In the absence of topography, the dynamics of the convection \textcolor{\revtxt}{is} controlled by the non-dimensional Rayleigh, Ekman, and Prandtl numbers, defined by $Ra = g \alpha \Delta T d^3/\nu \kappa$, $Ek = \nu/(\Omega d^2)$, and $Pr = \nu/\kappa$, respectively. 
 These first two parameters determine the relative influence of the buoyancy and Coriolis forces in the system, whereas the Prandtl number determines the ratio of the thermal diffusion time to the viscous diffusion time. 
 For simplicity we set $Pr=1$ in all simulations. 
 \textcolor{\revtxt}{We refer the reader to the Methods for a complete description of the mathematical system, but note that $d$ is the dimensional depth of the outer core ($\sim 2260$ km).
 }
 \textcolor{\revtxt}{An important output parameter is the Reynolds number, $Re = U d/\nu$, where $U$ is a typical flow speed.
 The outer core is thought to be turbulent ($Re\gg1$) and yet strongly constrained by rotation so that $Ek $ and the Rossby number, $Ro = EkRe,$ are much less than unity.
 We refer to this regime as rapidly rotating and note that the parameter $\Rat = Ek^{4/3} Ra$ controls the convective state \citep[e.g.][]{sC61,kJ12}. In the limit of rapid rotation, the critical Rayleigh number scales as $Ek^{-4/3};$  this implies that $Ra$ must be at least this large to generate convection. It then becomes convenient to characterize the convective dynamics in terms $\Rat,$ which is an order one quantity in the limit of vanishing $Ek$.
 }
 \textcolor{\revtxt}{Due to} computational limitations, numerical simulations cannot employ the same non-dimensional parameters of the core. 
 Rather, as is common practice \citep[e.g.][]{uC10,mC21}, our goal is to vary these parameters systematically to identify the appropriate asymptotic regime such that our findings can be extrapolated to core conditions. 
 For the topographic bump, we fix the location and width, and vary the amplitude, $\ep$. 
 \textcolor{\revtxt}
 { The viscous Ekman layer that is present along both the inner and outer boundaries has a thickness of $\de =O( Ek^{1/2})$. A conservative estimate for CMB topography is $1\text{ km}$, which suggests that $\epsilon /\de\sim 10^{4}$ in the core. To preserve the relevant dynamics, it is necessary to maintain the ordering of length scales $\epsilon \gg \de$. In our simulations the depth of the Ekman layers are greater than $10^{-3}$ so we employ values of $\ep$ that exceed this depth. Thus, three non-dimensional control parameters ($Ek, \Rat, \ep$) are varied in this investigation. }

 \textcolor{\revttxt}{In this study, over our investigated range of parameters, we confirm that topographic torques scale linearly with topographic amplitude. The simulation data also suggests a quadratic scaling with flow speeds, which can be explained via analysis of the governing equations. These findings indicate that small scale convective turbulence in the outer core is capable of generating topographic torques of sufficient magnitude to explain decadal and sub-decadal LOD variations.}
\section*{Results}

\begin{figure}
	\begin{center}
\subfloat[][]{\includegraphics[width=0.31\textwidth]{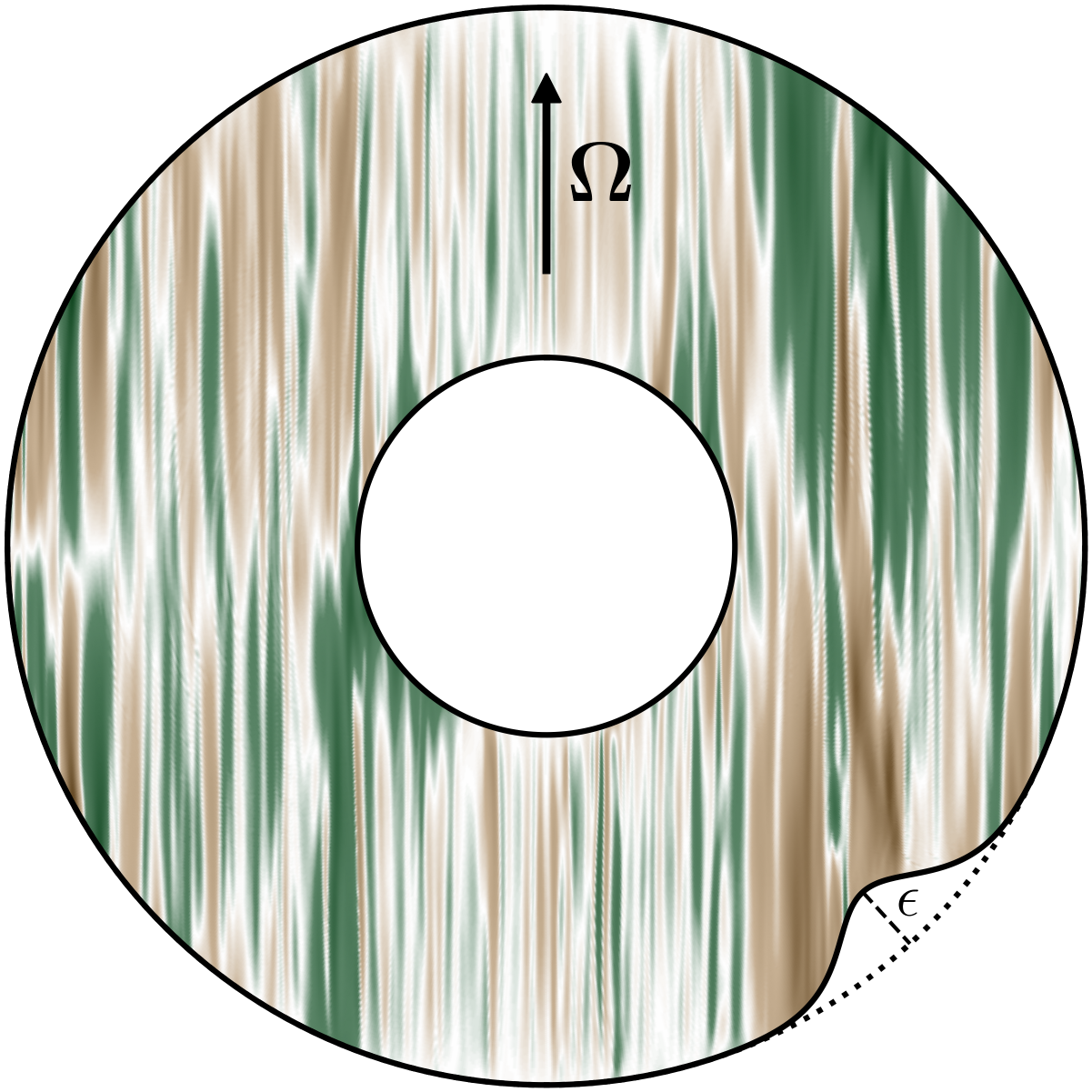}} \quad 
\subfloat[][]{\includegraphics[width=0.31\textwidth]{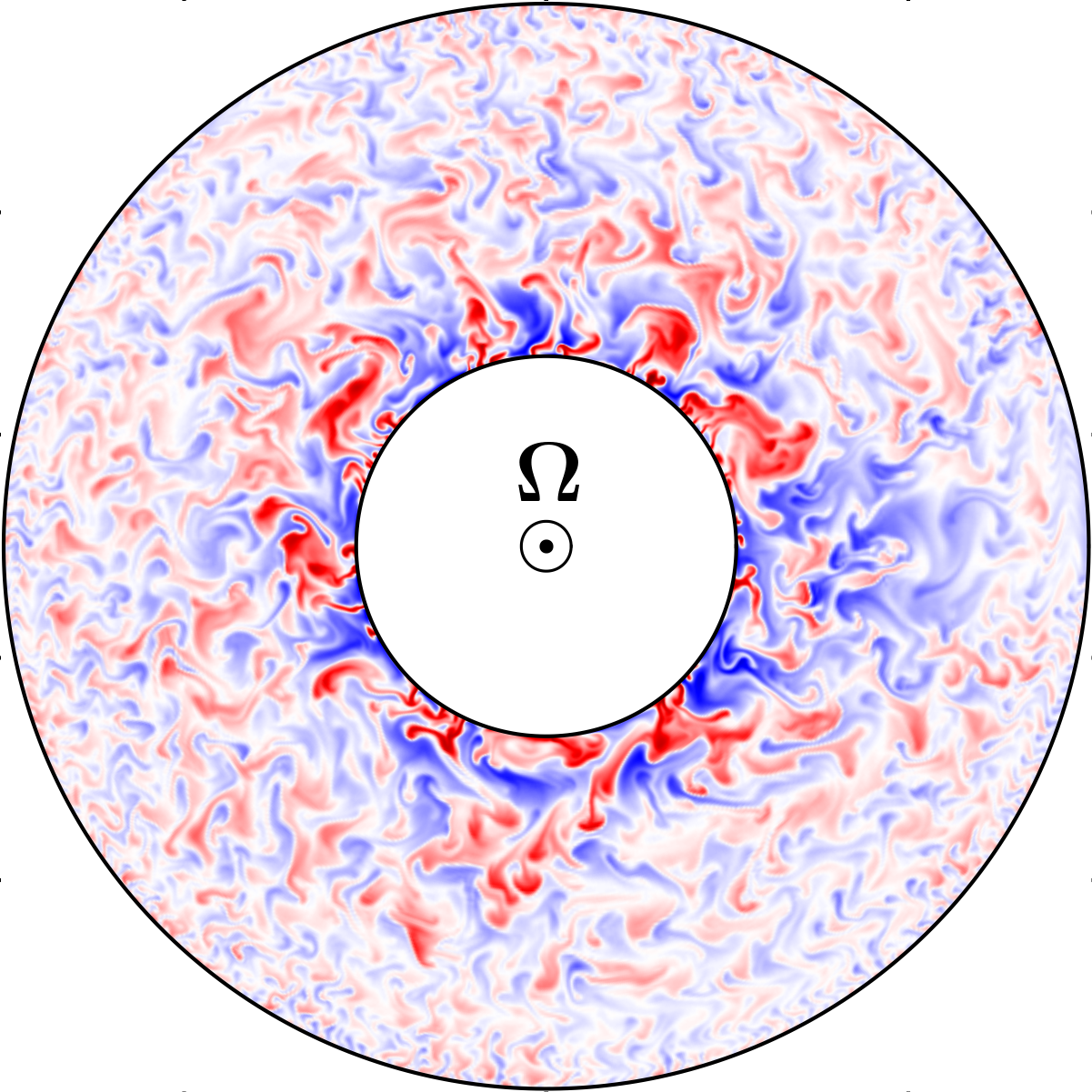}} \quad
\subfloat[][]{\includegraphics[width=0.31\textwidth]{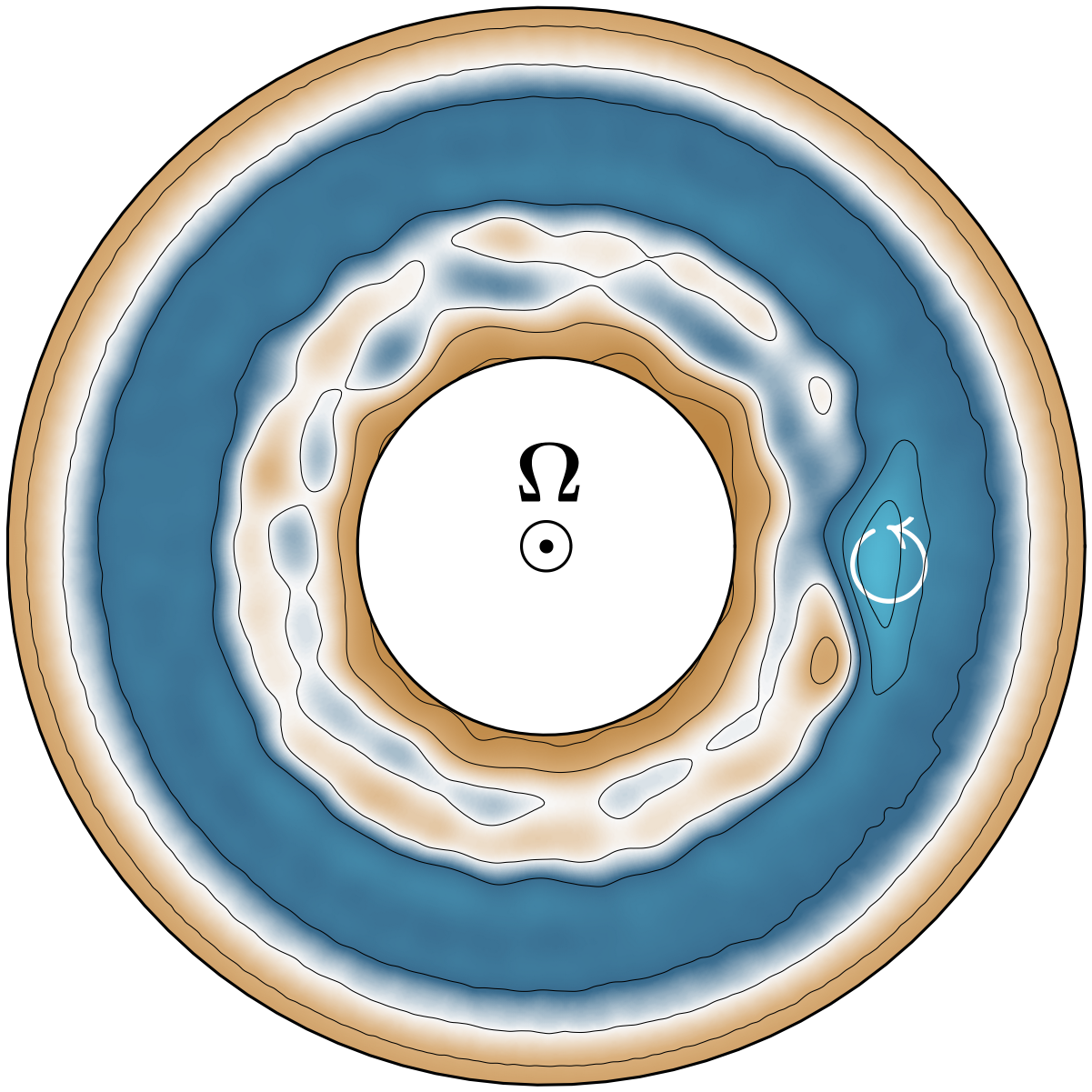}}
	\end{center}
	\caption{\textbf{Flow visualizations.} Visualizations for the most turbulent simulation: $Ek = 10^{-6}$, $\Rat = 60$, and $\epsilon  = 0.2$. (a) meridional cross-section of the instantaneous vertical velocity; (b) equatorial cross-section of the instantaneous temperature field with the conductive profile and zonal-average removed. (c) Time- and depth-averaged pressure. Blue (brown) corresponds to pressure lows (highs).
The circular arrow in (c) indicates the direction of the cyclonic flow over the bump.
	}
	\label{f:visu}
\end{figure}

Visualizations of the flow field are shown in Fig. \ref{f:visu} for the most extreme rotation rate ($Ek = 10^{-6}$) \textcolor{\revtxt}{at} $\Rat = 60$ and $\epsilon = 0.2$. 
\textcolor{\revtxt}{The meridional slice shows} the presence of \textcolor{\revtxt}{columnar structures} that effectively span the entire local depth of the domain. These structures drift with time and interact with the CMB topography. This view, along with equatorial slices, show the hallmark of turbulent rotating spherical convection: strong mixing in planes perpendicular to the rotation axis, yet approximately invariant parallel to the rotation axis. 
The dynamical effects of the topography are present in these snapshots of the flow\textcolor{\revtxt}{. In Fig. \ref{f:visu}(b)} we find subdued mixing in the column extending axially from the bump with locally cold fluid indicated by the predominantly blue color. 
Further evidence that this bump impacts the dynamics of the core is found in the time- and axially-averaged pressure field shown in Fig. \ref{f:visu}(c); a steady, cyclonic gyre is observed directly above the topography. 
%This field characterizes the large scale flows, with positive (negative) values indicating eastward (westward) flow. 

\begin{figure}
\centering
\subfloat[][]{\includegraphics[width=0.45\textwidth]{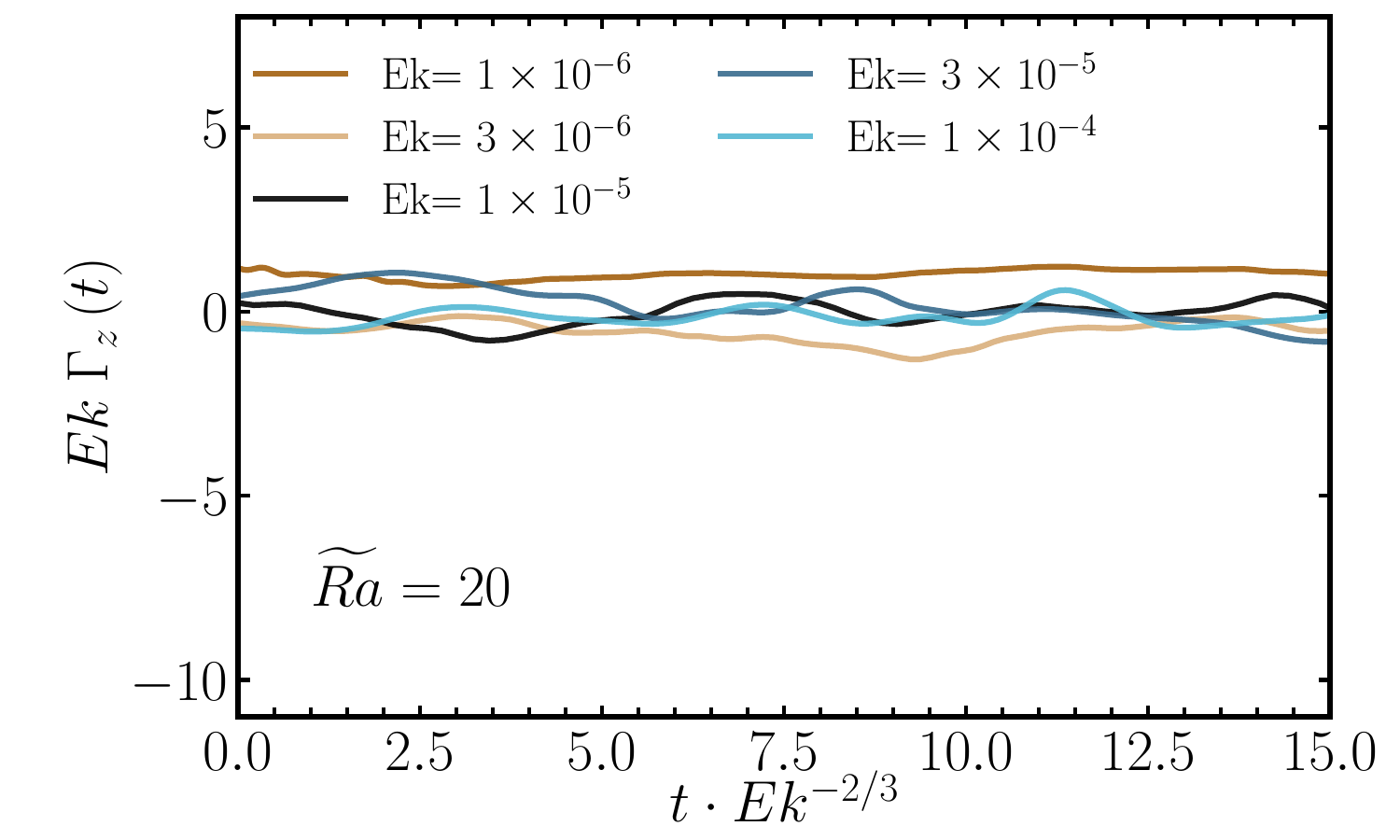}}
%\subfloat[][]{\includegraphics[width=0.32\textwidth]{figures/torque_ts_turb}}
\subfloat[][]{\includegraphics[width=0.45\textwidth]{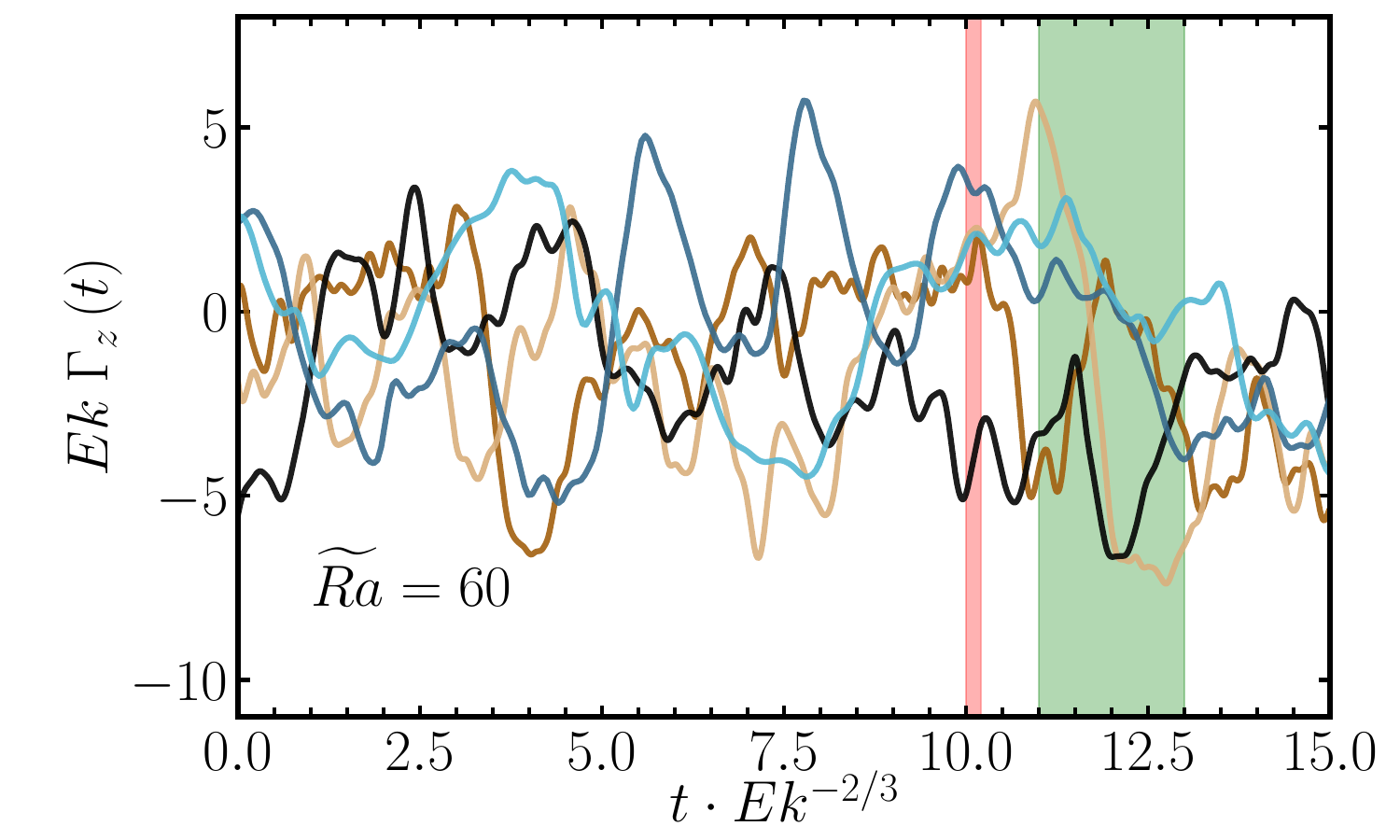}}\\
\subfloat[][]{\includegraphics[width=0.45\textwidth]{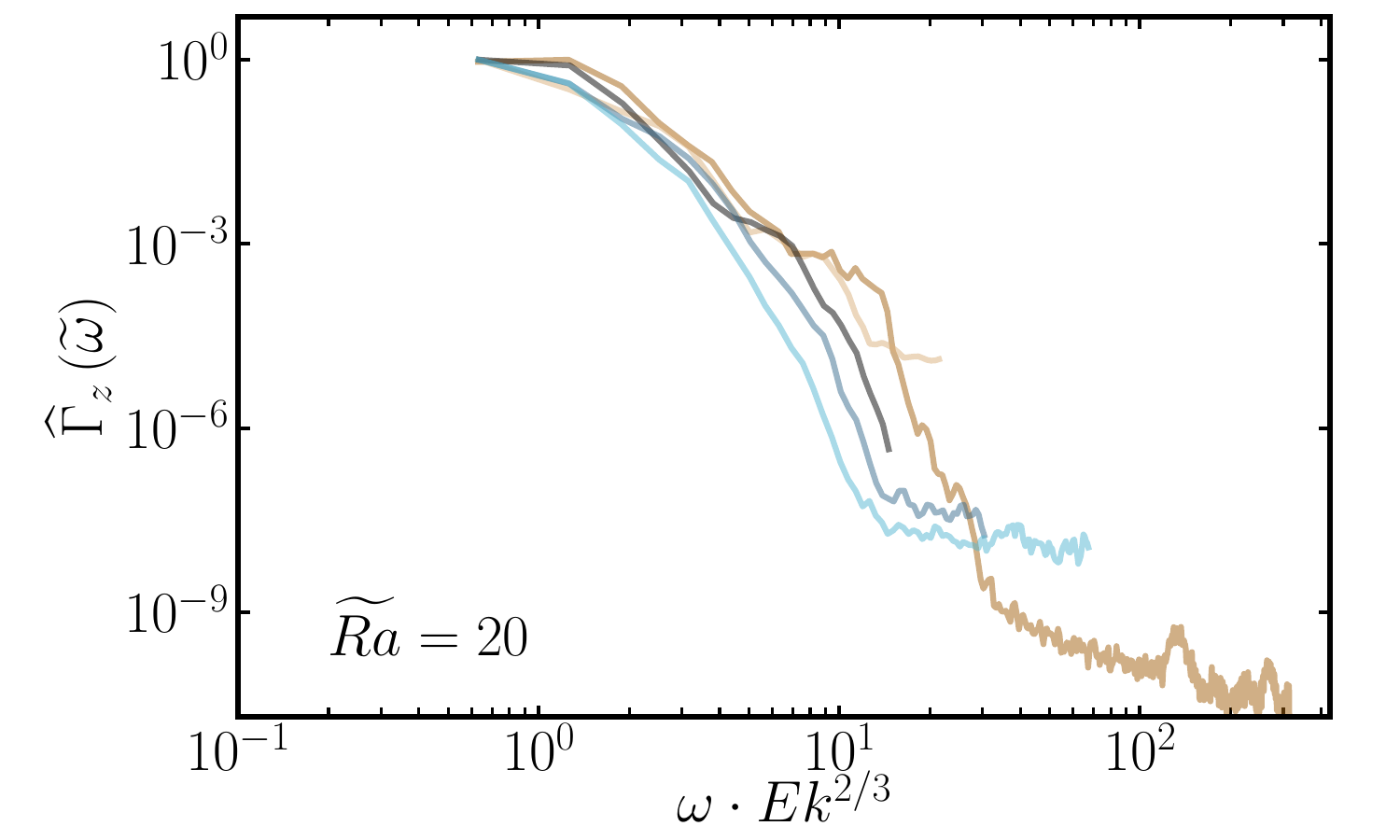}}
%\subfloat[][]{\includegraphics[width=0.32\textwidth]{figures/torque_pd_turb}}
\subfloat[][]{\includegraphics[width=0.45\textwidth]{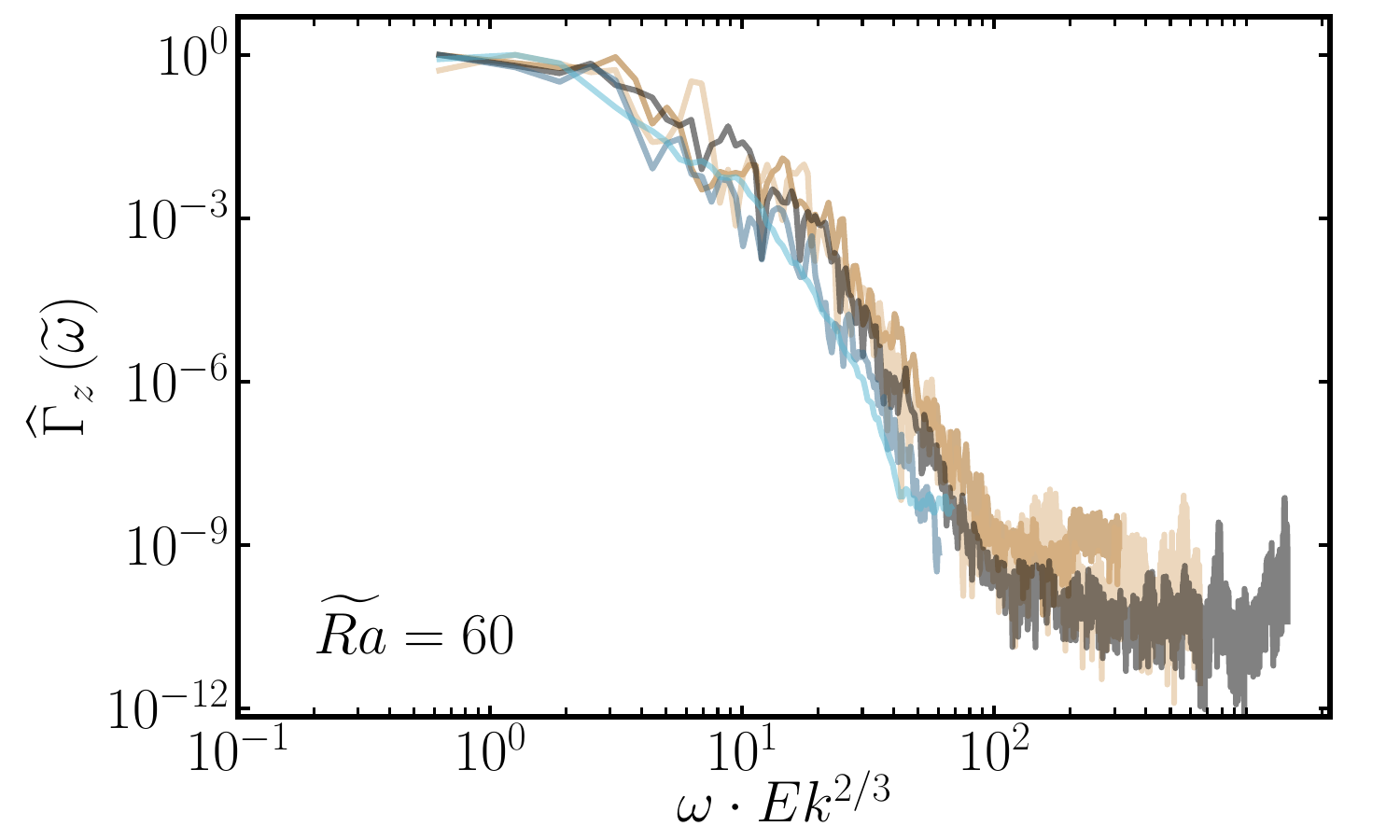}}
	\caption{\textbf{\textcolor{\revttxt}{Temporal variation of the reduced torques.}} (Top row) Time series of the reduced torque and (bottom row) the corresponding periodograms of the torque for $\epsilon  = 0.2$. Left $\Rat =20$ and right $\Rat= 60$ for all $Ek$. Red and green bars in (b) correspond to decadal and centenary timescales, respectively. 
%Periodograms were averaged over a Hann window of $\widetilde{t} = 10$ and normalized by the peak. 
}
\label{f:topo_ts}
\end{figure}
Time series and corresponding periodograms of the topographic torque are shown in Fig. \ref{f:topo_ts} for \textcolor{\revtxt}{$\epsilon = 0.2$, $\Rat = (20, 60)$ and all values of $Ek$.}
\textcolor{\revtxt}{The asymptotic theory of rapidly rotating convection\citep{kJ98a,mS06} predicts that the topographic torque should scale with the Ekman number as $Ek^{-1}$. Details on the origin of this scaling are provided in Methods D. We refer to the quantity $Ek \Gtz$ as the reduced torque, in analogy to $\Rat$, and plot it in Fig. \ref{f:topo_ts}(a-b). }
%\textcolor{blue}{Did something get deleted here? We are plotting the reduced torque before we define it?}

In the rapidly rotating regime, we expect the temporal dynamics to be dominated by the convective Rossby waves. In our non-dimensonal units, these waves have angular frequencies that vary as $\omega =O( Ek^{-2/3})$ \textcolor{\revtxt}{in the limit of vanishing Ekman number \citep[e.g.][]{cJ00}}. 
Thus, in order to meaningfully compare the simulations at different rotation rates \textcolor{\revtxt}{(Ekman numbers)}, the data is plotted against the rescaled time and frequency, $\tt = Ek^{-2/3} t$ and $\ot = Ek^{2/3} \omega$, respectively.
The periodograms are an estimate of the spectral density and are calculated with Welch's method \citep{pW67} over a Hann window of size $\textcolor{\revtxt}{\Delta}\widetilde{t} = 10.$ \textcolor{\revtxt}{The data are normalized by the power in the lowest frequency mode.}
Here we observe that in these rescaled units the simulations are characterized by similar time scales, showing that the asymptotic state relevant for understanding core dynamics is present in the models.
\textcolor{\revtxt}{
	Although this timescale is not commonly reported in previous studies of rotating convection and dynamos, we can rescale previously published temporal data into units of $\tt$. For instance, Schaeffer et al.\citep{nS17} studied the dynamics of rapidly rotating dynamos and report dominant frequencies in the range $10^{-4} - 10^{-3}$ when non-dimensionalized by the rotation rate; these frequencies are equivalent to $\tt = O(1) - O(10)$ and are therefore similar to the timescales in our simulations.
}
\textcolor{\revttxt}{
	Estimates for the convective overturn time based on the geomagnetic secular variation also suggest timescales of $\tt = O\lb 1\rb$\citep{cB16}.
}
As $\Rat$ is increased the amplitude of the fluctuations increase, as does the temporal  complexity. However the periodograms do not differ substantially between different $\Rat,$ suggesting the increased complexity at higher $\Rat$ is restricted to smaller amplitude fluctuations.

Under Earth-like conditions, a decadal timescale \textcolor{\revtxt}{($10$ years)} corresponds to \textcolor{\revtxt}{$\Delta\widetilde{t} \approx 0.2$}.
The red shaded region in Fig. \ref{f:topo_ts}(c) corresponds to \textcolor{\revtxt}{$\Delta\tt = 0.2$} while the green shaded region corresponds to \textcolor{\revtxt}{$\Delta\tt = 2.0$ ($100$ yrs.).}
Although our current data suggest that fluctuations take place on a timescale closer to $100$ years,
%Our current data suggests that fluctuations take place on a dimensional timescale closer to $100$ years,
the simulations are far less turbulent than the expected conditions of the core; accepted values of flow speeds for the core are at least five orders of magnitude greater than the simulated values. 
As the forcing is increased and the flow becomes more turbulent, the advective timescales are known to decrease\textcolor{\revtxt}{\citep{sP00}}, and this behavior is also observed in the simulations.
\textcolor{\revtxt}{Therefore, we believe that the timescales of the fluctuations make topographic torques good candidates for both the $\sim 6$ and $\sim 60$ year LOD fluctuations.
}
%Moreover, it is known that interpretation of LOD data is dependent on the filtering.  geomagnetic jerks suggest a strong correlation between outer core flow and the sub-decadal LOD variations \cite{rH13}.

\begin{figure}
\centering
\subfloat[][\label{f:tq_a}]{\includegraphics[width=0.48\textwidth]{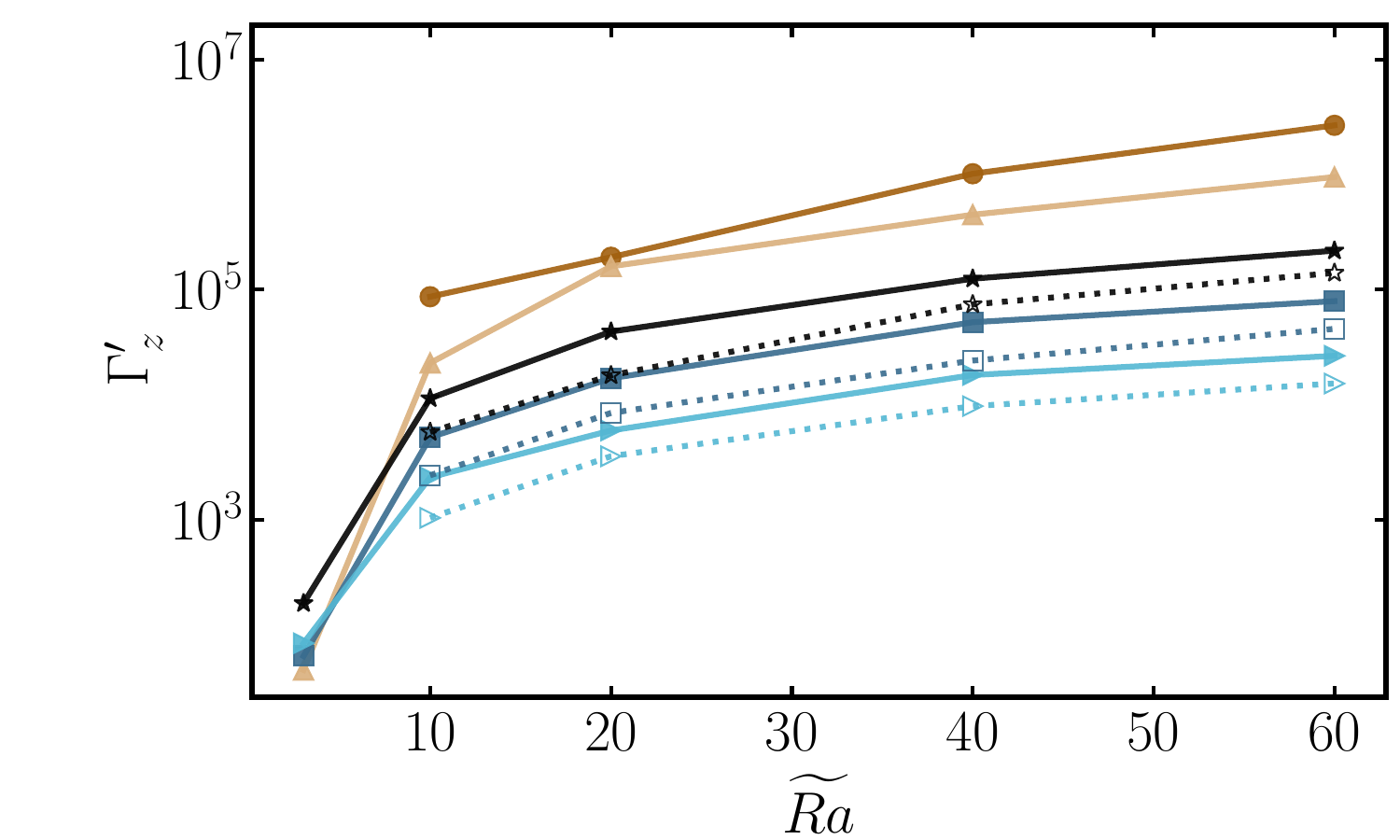}}
%\subfloat[][\label{f:tq_b}]{\includegraphics[width=0.48\textwidth]{figures/topo_std_Ek.png}} \\
\subfloat[][\label{f:tq_c}]{\includegraphics[width=0.48\textwidth]{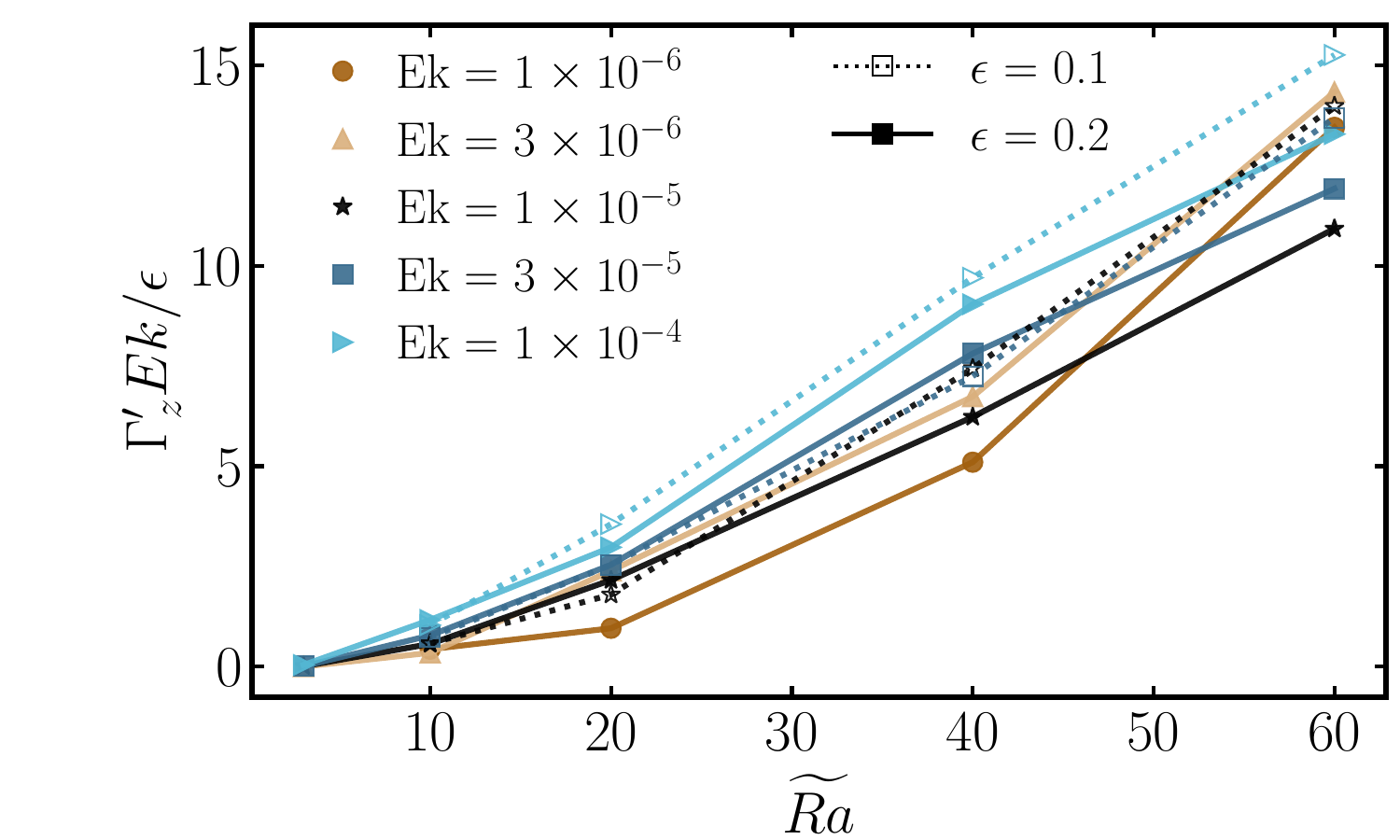}}
%\subfloat[][\label{f:tq_d}]{\includegraphics[width=0.48\textwidth]{figures/topo_std_Ekep2.png}}
	\caption{\textbf{Axial topographic torque versus reduced Rayleigh number.}  
	(a) Raw (unscaled) fluctuating axial torque and (b) \textcolor{\revtxt}{torque normalized by the predicted asymptotic scaling, $\ep/Ek$. \textcolor{\revttxt}{The reduced Rayleigh number is defined by $\Rat = RaEk^{4/3}$.} } 
	}
	\label{f:tp_torque}
\end{figure}
The standard deviation of the axial topographic torque ($\Gstd$) is computed for all simulations and plotted in Fig. \ref{f:tp_torque}(a). 
Dashed (solid) lines with open (solid) symbols correspond to $\ep = 0.1$ ($\ep = 0.2$). 
For a fixed topographic amplitude we observe an increase in $\Gstd$ as $Ek$ is reduced 
\textcolor{\revtxt}{
	at fixed $\Rat$},
	showing that CMB topography has a stronger influence on \textcolor{\revtxt}{super-critical} core flows as the rotation rate increases.
We emphasize that the abscissae are the reduced quantity $\Rat$. In dimensional terms this means that as the Coriolis force is increased, so too is the buoyancy force so that the flow remains super-critical. 
We also find that $\Gstd$ increases as $\Rat$ is increased, demonstrating that larger pressure gradients develop within the flow as the buoyancy force is increased.

\textcolor{\revtxt}{A plot of the reduced torque} is shown in the \textcolor{\revttxt}{Supplementary} Fig. \ref{f:tp_torque_supp}(a).
The reduced torque is order unity across the entire range of parameters investigated and is separated into two distinct branches corresponding to $\epsilon  = 0.1$ and $\epsilon  = 0.2$.
%Figure \ref{f:tp_torque}(b) shows that the reduced torque is order unity across the entire range of parameters investigated, providing strong support for the predicted asymptotic scaling. 
%We also see that, as might be expected, the reduced torque effectively separates into two distinct branches for each of the two amplitudes shown; i.e.~larger $\ep$ yield larger torques for a given value of $\Rat$. 
To determine how the torque depends on topographic amplitude we rescale the reduced torque by $\epsilon$ as shown in Fig. \ref{f:tp_torque}(b).
Good collapse of the data is observed \textcolor{\revtxt}{across the range of $\Rat$ investigated}.
Rescaling the torques by $Ek/\epsilon^{2}$, as shown \textcolor{\revttxt}{in Supplementary} Fig. \ref{f:tp_torque_supp}(b), yields values that are no longer order unity. 
\textcolor{\revtxt}{Furthermore,} there is \textcolor{\revttxt}{notable} separation in the data corresponding to different amplitudes which is is strongly suggestive that a scaling of $\Gstd =O(\ep^2)$ is not represented by our data.

\begin{figure}[H]
	\begin{center}
		\includegraphics[width=0.47\textwidth]{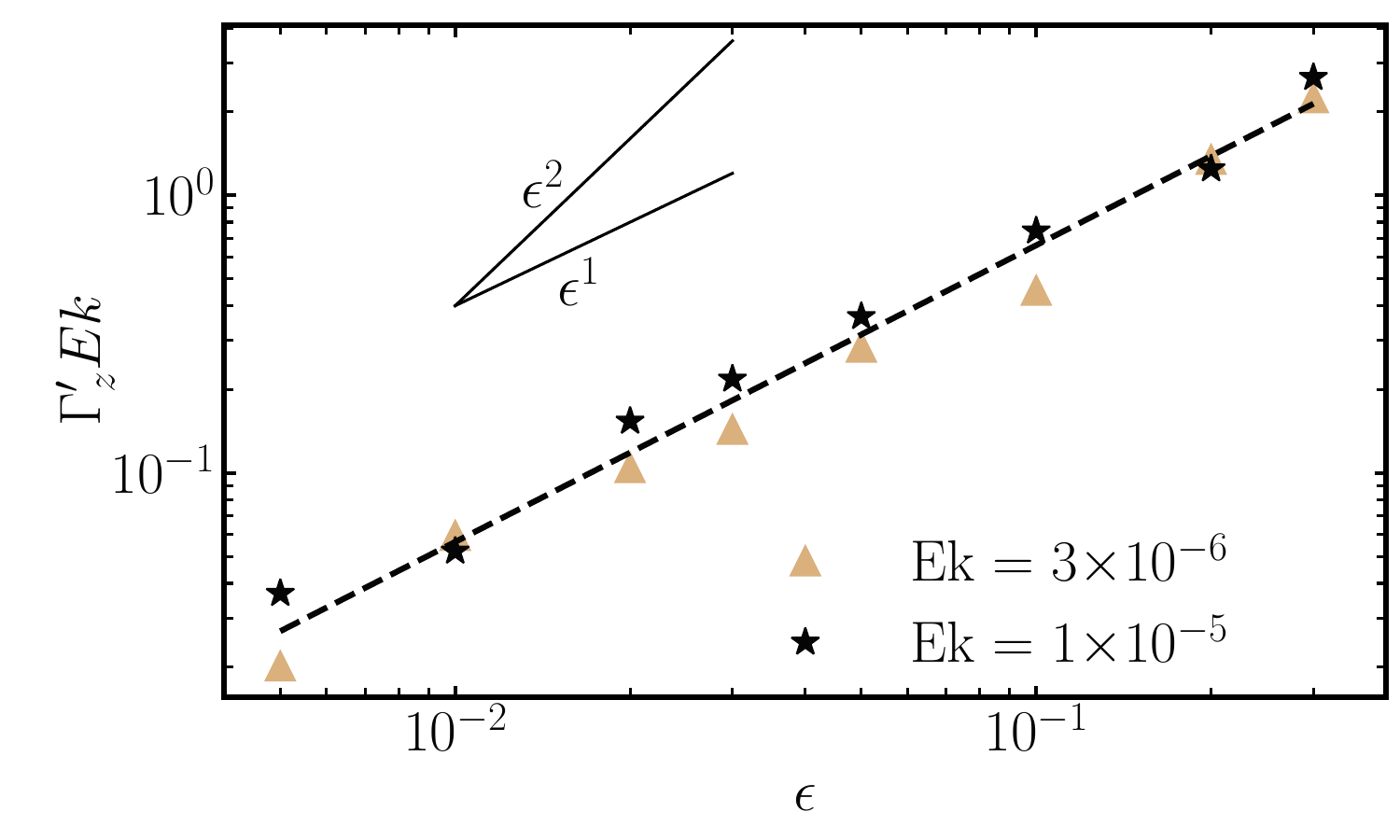}
        %\subfloat[][]{\includegraphics[width=0.48\textwidth]{figures/torque_std_vs_re_raw.png}}
	\end{center}
	\caption{\textbf{Reduced topographic torque versus topographic amplitude.} $\Rat = 40$ and $Ek = (10^{-5}, 3\times10^{-6})$. \textcolor{\revttxt}{The abscissa is $\epsilon$, the non-dimensional topographic amplitude.} A least squares fit of the data gives a scaling of $\Gstd Ek = 7.7 \epsilon^{1.07}$ shown with the dashed line. Linear and quadratic scalings in $\epsilon$ are shown for reference. 
	}
	\label{f:topo_sweep}
\end{figure}
Further testing of the amplitude dependence is carried out by varying $\ep$ for Ekman numbers $Ek = [10^{-5},3\times10^{-6}]$ and fixed $\Rat = 40$.
\textcolor{\revtxt}{Fig. \ref{f:topo_sweep}} shows the computed reduced torque for these cases. 
A least squares power law fit to the data yielded $\Gstd Ek = 7.7\epsilon^{1.07}$, and is shown by the dashed line. Slopes of $\ep$ and $\ep^2$ are shown for reference. These results provide further evidence that the topographic torque scales linearly with topographic amplitude, as would be expected if the primary component of the pressure that contributes to the torque is that associated with the convection (i.e.~the pressure variation is, to leading order, not dependent on $\ep$).

%\begin{figure}
%	\begin{center}
%	\subfloat[][]{\includegraphics[width=0.45\textwidth]{figures/topo_std_EkRa2.png}} \qquad
%        \subfloat[][]{\includegraphics[width=0.45\textwidth]{figures/torque_std_vs_re_raw.png}}
%	\end{center}
%	\caption{(a) Scaled topographic torque as a function of thermal forcing ($\Rat$).
%		(b) Topographic torque scaled by $Ek^{1/3}/\epsilon$ as a function of flow speeds ($Re$). The data appears to collapse well and looks to be approaching a $Re^{2}$ scaling for the largest $Re$ (lowest $Ek$). 
%	}
%	\label{f:topo_scaling}
%\end{figure}

\begin{figure}
	\begin{center}
\subfloat[][]{\includegraphics[width=0.47\textwidth]{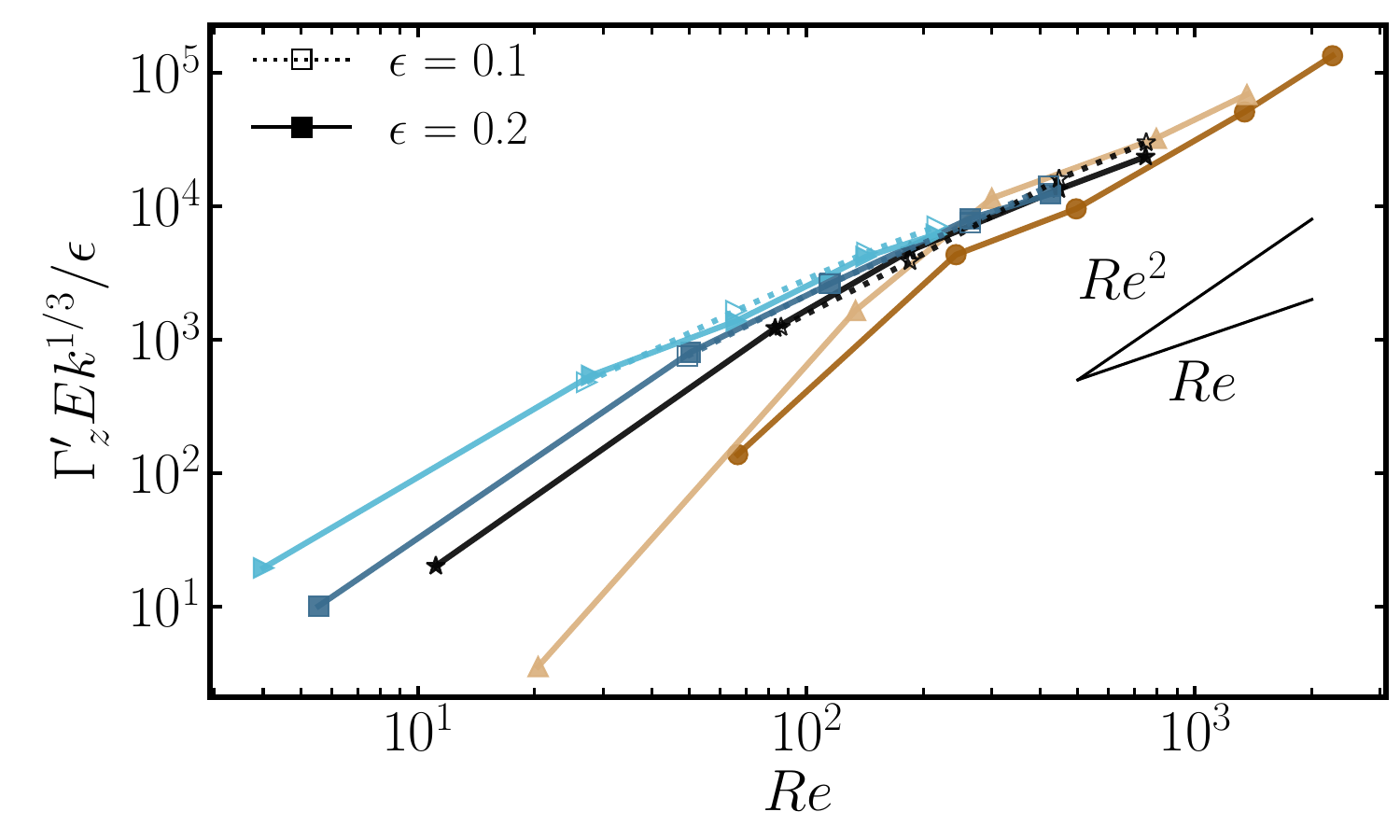}}
		\qquad
		\subfloat[][]{\includegraphics[width=0.47\textwidth]{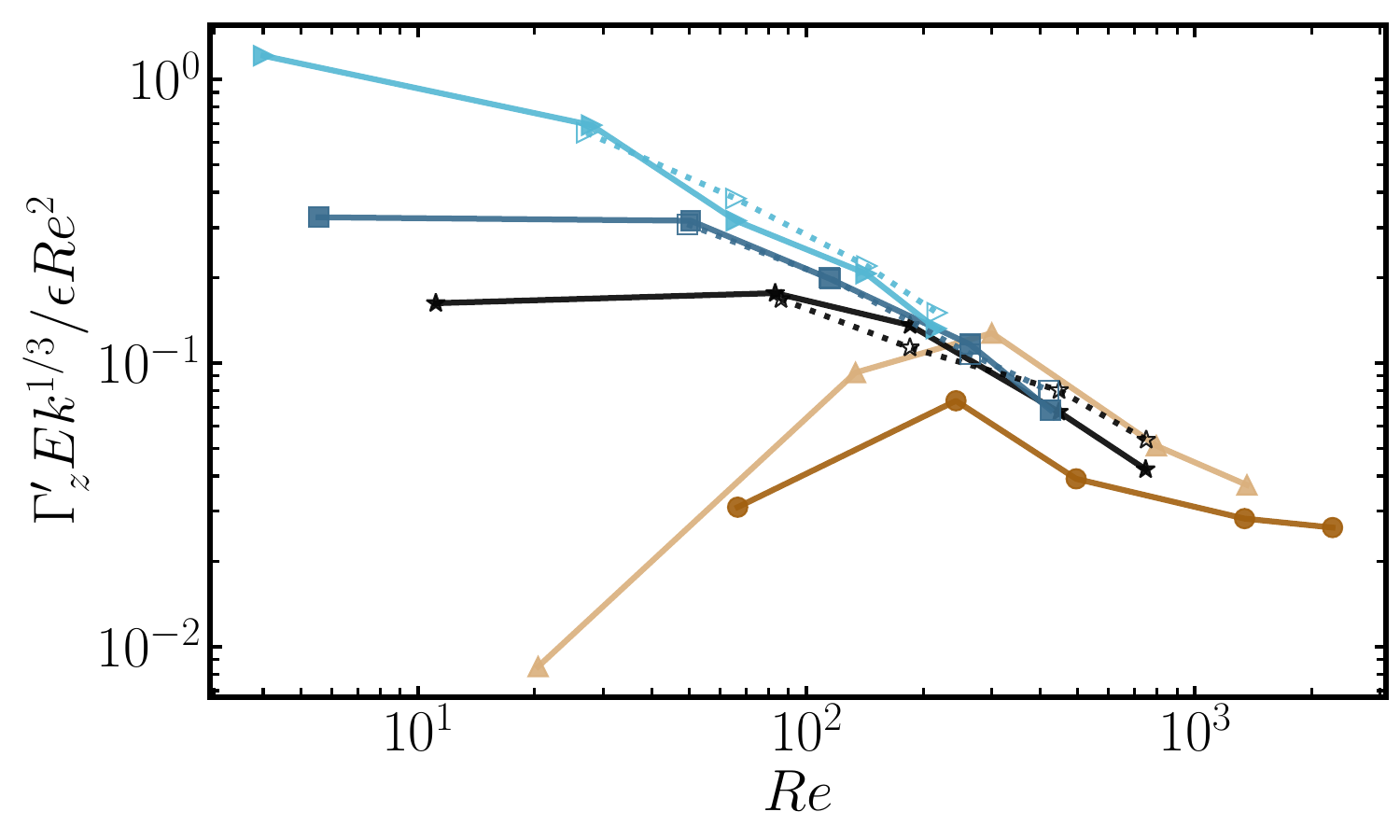}}
        %\subfloat[][]{\includegraphics[width=0.48\textwidth]{figures/torque_std_vs_re_raw.png}}
	\end{center}
	\caption{\textcolor{\revttxt}{\textbf{Dependence of topographic torque on flow speed.}}
		(a) Topographic torque scaled by $Ek^{1/3}/\epsilon$ as a function of flow speed ($Re$); \textcolor{\revtxt}{slopes of $Re$ and $Re^2$ are shown for reference. (b) Compensated topographic torque versus $Re$. Symbols are as defined in Fig. \ref{f:tp_torque}.}
	}
	\label{f:topo_scale}
\end{figure}
To extrapolate our findings to the conditions of the core, we require the functional dependence $\Gstd = f(\Rat)$, or, alternatively, $\Gstd = g(Re)$, for some functions $f$ and $g$. 
\textcolor{\revtxt}{The asymptotic theory presented in Methods D predicts that for $Ek \rightarrow 0$, and sufficiently large $\Rat$, the torque should scale as 
\be
\Gstd = O \lb \epsilon Re^{2} Ek^{-1/3} \rb.
\label{e:tq_full_scale}
\ee
This relationship arises from an approximate balance between the ageostrophic component of the pressure gradient force and advective nonlinearities in the prognostic evolution equation. 
To test this scaling we plot the quantity $\Gstd Ek^{1/3}/\epsilon$ as a function of $Re$ in Fig. \ref{f:topo_scale}(a). 
With the exception of the simulations with small values of $\Rat$ in which the flow is very close to the onset of convection and is therefore dominated by few oscillatory modes, the $Ek^{-1/3}\epsilon$ scaling collapses the data well. Only our most extreme simulations are approaching the quadratic flow speed scaling ($Re^{2}$), as indicated in Fig. \ref{f:topo_scale}(b) which shows the fully compensated torque, $\Gstd Ek^{1/3}/(\epsilon Re^2)$. 
Ultimately we want to be able to predict torques using only the input parameters, rather than an output parameter such as $Re$. Previous work \citep{sM21, tO23, jN24} has found that flow speeds scale approximately as $Re \sim Ra Ek Pr^{-1} = \Rat Pr^{-1} Ek^{-1/3}$ when $\Rat$ becomes large. Omitting the Prandtl number dependence, this leads to $\Gstd \sim \ep \Rat^2 Ek^{-1}$. \textcolor{\revttxt}{Supplementary} Fig. \ref{f:topo_Ra2} shows the reduced torque rescaled by $\epsilon \Rat^{2}$ and a reasonable collapse of the data is observed indicating this dependence may be robust at large Rayleigh numbers.
}

%\textcolor{\revtxt}{Figure \ref{f:tp_torque}(b) shows that the $\Rat$-dependence for the torque becomes stronger as the Ekman number is reduced. Further evidence for this enhancement with decreasing $Ek$ is shown in Extended Data Figure \ref{f:topo_Ra2} where the reduced torque is rescaled by $\epsilon \Rat^{2}$ and a reasonable collapse of the data is observed.}

\section*{Discussion}

The significance that topographic torques play in explaining the \textcolor{\revtxt}{large amplitude} fluctuations in the LOD has been debated for more than 50 years \citep{rH69}. This debate, which is centered on two competing hypotheses, has focused on how the torques depend on the typical amplitude of the topography.
It is known that a non-axisymmetric pressure gradient is required to generate non-trivial axial torques on the mantle. The first hypothesis assumes that the flow in the core is predominantly axisymmetric and the topography induces the necessary non-axisymmetric pressure gradients \citep[e.g.][]{aA77b}. Models using this hypothesis yield topographic torques that scale quadratically with the topographic amplitude. 
The second hypothesis considers the convection to be the dominant source of asymmetry such that the pressure gradients do not depend on topographic amplitude to leading order, which leads to a linear scaling. 
Our investigation has shown that this second mechanism dominates when convection is considered in the problem -- convection-driven waves with angular frequency $\omega=O( Ek^{-2/3})$, which are inherently non-axisymmetric, yield pressure gradients that interact with the topography to produce a torque that is linearly proportional to the topographic amplitude. 
%\textcolor{\revtxt}{Perturbations to the axisymmetric flow generated by the topography are still geostrophic to leading order, and therefore will not contribute to the torques.
%}

The results also show that the topographic torque scales nearly quadratically with the flow speeds at the most extreme parameter values accessed in the simulations. 
This scaling is nearly consistent with the classical picture of `form drag' \textcolor{\revtxt}{in which a flow traveling with speed $U$ is diverted by an obstacle.  In turbulent flows, the pressure near a boundary is expected to scale quadratically with flow speed in order to allow for complete stagnation of the incident flow\citep{gB00}. However, in this study rotation appears to play an additional role via the factor of  $Ek^{-1/3}$.} 
This Ekman number dependence suggests that torque is dependent not only on flow speeds, but also the spatial structure of the underlying convection, which is controlled by the importance of rotation through the Ekman number. 
\textcolor{\revtxt}{We stress that the present study was performed at fixed topographic width, and we expect different topographic shapes to influence the torque (see equation \ref{E:torque}).}

We present a scaling argument based on quasi-geostrophic theory \cite{mS06,mC13} which recovers the empirical scaling for the topographic torque given by $\Gstd =O( \epsilon  Re^{2}Ek^{-1/3})$.
%Thus, our investigation shows that an approximate scaling of the topographic torque is given by $\Gstd \sim \ep Re^2 Ek^{-1/3}$. 
It is \textcolor{\revttxt}{notable} that this scaling differs from the $Re/Ek$ scaling originally predicted by Hide \citep{rH69}. 
	Hide used the geostrophic scaling (equation \ref{E:geo}) to predict the magnitude of $\Gtz$, although he noted that there would be partial cancellation in the surface integral. However, Anufriev and Braginsky\citep{aA77b} pointed out that there is total cancellation if only the geotrophic flow is considered. An important consideration is that this argument holds for all geostrophic structures. 
	Therefore, topographic torques arise only from scales where a mechanism to disrupt geostrophy, such as viscosity or magnetism, are \textcolor{\revttxt}{present}. Our interpretation and data supports this conclusion, implying that perturbations to the geostrophic pressure need to be considered. 

\textcolor{\revtxt}{We report our scaling for $\Gstd$ in terms of $Re$ and $Ek$ rather than the control parameters $Ra,Ek$ and $Pr$ because the data and asymptotic theory are more directly supportive of the former. Regardless,} our derived torque scaling should be understood to depend on all of the control parameters. 
We note that the flow speed, and therefore the Reynolds number, is not independent of $Ek$. For instance it is known that when the system is strongly forced by fixed temperature boundary conditions, the Reynolds number scales approximately like $Re =O( Ek Ra/Pr)$ and the singularity as $Ek\to 0$ is avoided\citep{tO23}. 

\textcolor{\revttxt}{A dimensional torque of nearly $10^{18}$ N$\cdot$m is required to explain observed LOD variations\citep{pR12}}. In dimensional form our scaling becomes 
\begin{equation}	
\Gstd_{\text{, dim}} \sim \epsilon Ek^{-1/3}  \rho   U^{2}d^{3},
	\label{e:Gamma_dim}
\end{equation}
where $U$ is a characteristic flow speed and fluid density is denoted by $\rho$. 
Using estimates for the outer core we have $\rho  \sim 10^{4}$kg m$^{-3}$, $\epsilon \sim 10^{-3}$ (corresponding to an amplitude of $2.26$ km), $U = (1-5)\times 10^{-4}$ m s$^{-1}$,  $d = 2260$ km, and $Ek = 10^{-15}$. These values give a dimensional torque of $(1.2\times 10^{17}-2.9\times 10^{18})$ N$\cdot$m, which is consistent with the required torques.

%\textcolor{\revttxt}{Equation \ref{e:tq_full_scale} and the theory provided in Methods D and E require a choice for the scaling of the horizontal length scale $\ell$. In this work we have used the viscous $\ell= O(Ek^{1/3})$ scaling which arises from linear theory and has been shown to hold into the turbulent regime \citep{tO23}. However, the so called inertial scaling $\ell = O(Re^{1/2}Ek^{1/2})$ is proposed and supported by simulations as well \citep{cG19,cG25}. 
%	It is believed that the currently accessible parameter regime in both experiment and simulation is unable to distinguish between these two predictions \citep{jA24}, however under the conditions of the Earth's core the associated length scales are expected to differ by two orders of magnitude\citep{cG25}. Adopting the inertial scale would decrease the predictions for the torque by a factor of $10^{2}$ and the relevant timescale by a factor of $10^{4}.$ }
\textcolor{\revttxt}{Equation \ref{e:tq_full_scale} and the theory provided in Methods D and E require a choice for the scaling of the horizontal length scale $\ell$. In this work we have collapsed the numerical data using the asymptotic scale $\ell= O(Ek^{1/3})$, which is well known from linear theory, but also holds in the strongly turbulent regime \citep{mS06}. Some studies suggest an inertial scaling, characterized by $\ell = O\lb Ro^{1/2}\rb$, holds for the most energetic scale in the system \citep{cG19,cG25}. The most extreme simulations of rotating convective turbulence carried out to date do not show clear evidence of the inertial scaling \citep{tO23}, although it has been shown that these two theories are likely indistinguishable in the currently accessible parameter regime\citep{jA24}. This is likely attributable to the fact that both scalings can be shown to contain an $Ek^{1/3}$ factor \citep{jmA20, sM21,jN24}, and since the effect of particular flow scales on topographic torques is not known, we prefer the use of the $Ek^{1/3}$ scale which collapses all convective length scales.
}

%
%\textcolor{\revtxt}{ 
%	Considering equation (\ref{e:top_torque}), it may seem inconsistent that the recovered dimensional torques are similar to the previous predictions which used the asymptotically larger geostrophic pressure. We reconcile this by noting that the extra factor of $Re$ nearly covers the loss of the factor of $Ek^{-2/3}.$ 
%} \textcolor{blue}{Do we really need to mention this?}

\textcolor{\revtxt}{Several additional physical effects will be investigated in future studies to develop better understanding of topographic torques. For instance, magnetism and global scale topography likely also play a role in controlling LOD. }
However, because the primary flow quantities, such as flow speeds, show similar scaling behavior in simulations of both dynamos and convection \citep{eK13b,mC21,jN24}, the scaling of the topographic torques are unlikely to change fundamentally when magnetism is introduced.  
\textcolor{\revtxt}{We expect the CMB to be characterized by a broad range of topographic wavelengths\citep{pK21,aE18}, and the present work focused on a topographic bump with a wavelength that is small in comparison with the depth of the outer core. Thus, systematically varying the topographic wavelength is necessary to elucidate additional interactions between core flows and the CMB.  Earth's core is liquid metal and is therefore characterized by a thermal Prandtl number, $Pr$, that is less than unity. Our study is restricted to $Pr =1$, although it is known that convective dynamics are strongly influenced by the Prandtl number\citep{mC12b,jmA18,tV18b}. Zonal flows become more prominent \textcolor{\revttxt}{as} $Pr$ is reduced below unity\citep{jA01} -- understanding how this change influences topographic torques is of \textcolor{\revttxt}{considerable} interest for future investigations. For instance, it is known that when $Pr<1$ large amplitude standing Rossby waves can be excited by topography, though their influence on the topographic torque is still uknown\citep{mC12}. The possibility of stably stratified layers regionalized near the LLSVP's has been proposed by previous work, for instance see Mound et al.\citep{jM19}. The existence of such regions would likely have a direct impact on the scaling of topographic torques.}

\section*{Methods}
\renewcommand{\thetable}{\arabic{table}}

\subsection{Governing Equations}
For simplicity, the present work focuses solely on the hydrodynamic problem of convection of a Boussinesq fluid contained within a rotating spherical shell with inner radius $\ricb$ and outer radius $\rcmb$. In the absence of topography the aspect ratio of the shell is fixed at $\chi = \ricb/\rcmb = 0.35$. As discussed below, finite amplitude topography is superposed onto the outer shell. The physical properties of the fluid are the kinematic viscosity $\nu$, thermal diffusivity $\kappa$ and thermal expansion coefficient $\alpha$. 
The inner and outer boundaries are held at fixed temperatures $\ticb$ and $\tcmb$, respectively. 
$\Omega $ is the angular rotation rate.
We assume a uniform distribution of mass within the shell such that the gravitational acceleration varies linearly with radius and has magnitude $g$ at the outer boundary.
The governing equations reflect the conservation of momentum, energy, and mass and are non-dimensionalized by the shell gap-width $d = 2260\text{km}$, viscous diffusion time $d^{2}/\nu$, and temperature difference $\Delta T = \ticb - \tcmb.$ The equations then become
\begin{subequations}
	\begin{equation}
		\frac{\partial \mathbf{u}}{\partial t} + \mathbf{u}\cdot\nabla \mathbf{u} = -\nabla p +\nabla^{2}\mathbf{u} - \frac{2}{Ek}\mathbf{\hat{z}}\times\mathbf{u} + \frac{Ra}{Pr}\frac{\mathbf{r}}{\rcmb}\vartheta
		\label{e:ns}
\end{equation}
\begin{equation}
	\frac{\partial \vartheta}{\partial t} + \mathbf{u}\cdot \nabla \vartheta = \frac{1}{Pr}\nabla^{2}\vartheta	
	\label{e:energy}
\end{equation}
\begin{equation}
	\nabla\cdot\mathbf{u}  = 0	.
	\label{e:continuity}
\end{equation}
\label{e:govn_eqn}
\end{subequations}
The fluid variables are the velocity $\ub$, pressure $p$ and temperature $\vartheta$. The non-dimensional control parameters are the Rayleigh number, Ekman number, and the thermal Prandtl number, defined by, respectively,
\begin{equation}
	Ra = \frac{g \alpha\Delta T d^{3}}{\nu \kappa}, \qquad
	Ek = \frac{\nu}{\Omega d^2}, \qquad
	Pr = \frac{\nu}{\kappa}. \qquad
	\label{e:non_dim}
\end{equation}
The Prandtl number is fixed to unity. No-slip, impenetrable and isothermal boundary conditions are employed on the inner and outer surfaces. 
The Reynolds number is a diagnostic parameter that characterizes the flow speed. We define it as 
\begin{equation}
	Re = \overline{\sqrt{\left<\mathbf{u}\cdot\mathbf{u}\right>}},
	\label{e:re}
\end{equation}
where the overline denotes a time average and  $\left<\cdot\right>$ indicates a volume average.

\subsection{Model Setup and Numerical Details}
\begin{table}
\textcolor{\revtxt}{
\begin{center}
\begin{tabular}{|c|c|c|c|}
\hline
$\epsilon$ &Amplitude (km)&Aspect Ratio ($\epsilon /w$) &$\cos^{-1}(\mathbf{\hat{r}}\cdot\mathbf{\hat{n}})$\\
\hline
$0.005$&11.3 &$4.8\times 10^{-2}$ &$1.01^{\circ}$\\
\hline
$0.01$&22.6  & $9.5\times 10^{-2}$&$2.02^{\circ}$\\
\hline
$0.02$&45.2  &$0.019$ &$4.06^{\circ}$\\
\hline
$0.03$&67.8  &0.029 &$6.10^{\circ}$\\
\hline
$0.05$&113   & 0.048&$10.2^{\circ}$\\
\hline
$0.1$&226    &0.095 &$20.1^{\circ}$\\
\hline
$0.2$&452    & 0.19&$37.5^{\circ}$\\
\hline
$0.3$&678    & 0.29&$50.3^{\circ}$\\
\hline
\end{tabular}
\end{center}
}
\caption{
	\textbf{Topographic parameters.}
Geometric parameters of the topographic bumps employed in the simulations. Amplitudes scaled to outer core dimensions are provided for reference. \textcolor{\revttxt}{The non-dimensional amplitude is $\epsilon .$ }
The width of the bump, denoted by $w$, is fixed in the present study and is defined as the full width at $1\%$ maximum of the topography. 
The maximum angle between the surface normal vector $\mathbf{\hat{n}}$ and the radial unit vector $\hat{\mathbf{r}}$ is given in the last column. 
}
\label{t:bump_geo}
\end{table}

Given that the CMB is poorly constrained by seismology, we are motivated to use as simple a model as possible for its shape, whilst retaining the ability to easily control essential physical characteristics such as amplitude and width. 
In this regard we superpose a Gaussian bump on the outer boundary, as shown in Supplementary Fig. \ref{f:topo}. 
Rotating convection in a spherical shell is known to be anisotropic and inhomogeneous, which suggests the location of the topography may also play a key role, at least insofar as the dynamical consequences of the topography on the convective flows are concerned. 
For the present investigation, we place the bump in the lower hemisphere at a latitude that is intended to ensure a robust interaction with time-dependent convective Rossby waves and the geostrophic, zonal flows that develop in a rotating, spherical geometry.
We fix the coordinates of the center of the base of the bump to \textcolor{\revtxt}{ $\lb \theta_{0},\phi_{0}\rb $} such that it lies directly beneath the halfway point between inner and outer boundaries, as shown in Supplementary Fig. \ref{f:topo}(b). The spherical coordinates are thus $\lb \theta_{0},\phi_{0}\rb  = \lb 2.40,0\rb $.
The topographic shape function is given by
\begin{equation}
	h(\theta, \phi) =  \exp \left\{ -\frac{1}{\beta}
	\ls\frac{ \lb \mathbf{r} - \mathbf{r}_0\rb^{2}}{r_{cmb}^{2}}
	\rs
	\right\},
	\label{e:h}
\end{equation}
where $\mathbf{r}_{0}$ is the vector pointing from the origin to $\lb \rcmb,\theta_{0},\phi_{0}\rb $. 
The topographic amplitude is $\epsilon$ and $\beta$ denotes a measure of the width (lateral extent) of the topography. We fix the width in this study such that $\beta^{-1} = 40.$
As mentioned in previous work \citep{mC12}, the Gaussian profile is particularly useful since its representation in spectral space requires a broad range of spherical harmonics, and therefore lends the possibility of richer dynamical interactions with the convective flow field. 
\textcolor{\revtxt}{
	Table \ref{t:bump_geo} lists topographic parameters, including the aspect ratio for the topography, $\epsilon /w$, where $w$ is defined as the full width at $1\%$ maximum. Note that only the topographic amplitude is varied in the present study.}

The simulations are performed with the computational fluid dynamics code \texttt{Nek5000} \citep{nek5000}, which uses the spectral element method \citep{tP84} and therefore easily allows for solving the governing equations in domains with irregular boundaries. 
\texttt{Nek5000} has been extensively benchmarked and scales efficiently to large numbers of processors. 
The simulations were run on up to 2048 processors for our most demanding simulations. \texttt{Nek5000} uses a CFL condition to dynamically update the timestep. Our tests indicate that for identical non-dimensional control parameters the \texttt{Nek5000} code is approximately a factor of ten slower in comparison to the commonly used spectral code, \texttt{Rayleigh} \citep{hM16}, although spectral codes are generally incapabale of handling irregular boundaries.
All simulations used a degree $7$ polynomial interpolant within the spectral elements.
The most demanding simulations required 331,776 spectral elements, corresponding to almost $1.7\times10^8$ physical grid points. 
\textcolor{\revtxt}{The code was benchmarked against the hydrodynamic case 0 presented in (Christensen et al.)\citep{uC01} for a spherical geometry and all values were found to be within the recommended ranges. 
To our knowledge, no benchmark of convection in a deformed sphere exists, however extensive tests were performed to ensure that the expected boundaries were being implemented.}

\textcolor{\revtxt}{
The use of no-slip mechanical boundary conditions implies viscous torques act along both the inner and outer surfaces. Viscous torques were computed both as a means to check consistency in the numerical method and to compare with the topographic torque.
	The total viscous torque acting on both the inner boundary ($icb$) and the outer boundary ($cmb$) is calculated as
	\begin{equation}
	\Gtz^{v} = -\oint_{icb,cmb} \mathbf{r} \times \left[ \lb\nabla \ub +\lb\nabla\ub\rb^{T}\rb \cdot d\mathbf{a}\right] ,
		\label{e:vtq}
	\end{equation}
where the superscript $T$ denotes a transpose. \textcolor{\revttxt}{Supplementary} Fig. \ref{f:visc_torque_supp}(a) shows a representative time series of the pressure torque on the outer boundary and the sum of the viscous torques on the inner and outer boundary, as well as the rate of change of the axial angular momentum of the fluid $dL_z/dt$. We find excellent agreement between $-dL_z/dt$ and $\Gtz^{v}+\Gtz$, providing a verification of the calculation of the torques.}

The ratio of the standard deviations of the viscous and topographic torque on the outer boundary, denoted by $\gamma$, is shown in \textcolor{\revttxt}{Supplementary} Fig. \ref{f:visc_torque_supp}(b). This ratio is near unity near the onset of convection, but is generally a decreasing function of $\Rat$. These trends indicate that the viscous torques are likely \textcolor{\revttxt}{negligible} in the parameter regime of the core.

\subsection{Choice of parameter values}

Core turbulence is thought to be strongly constrained by the effects of the Coriolis force. 
Thus, our present goal is to model convection in this rotationally constrained regime while simultaneously allowing for as broad a range of spatiotemporal scales as is numerically feasible. 
Moreover, as is often done in studies of core dynamics, it is important to understand how the dynamics of the system change as the parameters are made more extreme. 
We therefore cover a broad range in Rayleigh number and Ekman number in an effort to elucidate the scaling behavior of the dynamics. 
Our simulations use Ekman numbers of $Ek = [10^{-6}, 3 \times 10^{-6}, 10^{-5}, 3 \times 10^{-5}, 10^{-4}]$ and reduced Rayleigh numbers of $\Rat = [3, 10, 20, 40, 60]$. 

Concerning the choice of topographic parameters, the present work is focused predominantly on how the topographic torque and associated convective dynamics depend on the amplitude, $\epsilon$. 
Although seismic studies and mantle convection simulations indicate that CMB topography amplitudes are unlikely to exceed $\epsilon = O(10^{-3})$, it makes little sense to use amplitudes this small given that the investigated range of Ekman numbers in our simulations is so distant from the Earth's core.
\textcolor{\revtxt}{The topographic amplitude at the CMB is thought to be \textcolor{\revttxt}{much} larger than the thickness of the Ekman boundary layer that is present. 
%For instance, assuming a topographic amplitude of $1$km so that $\ep \sim 10^{-4}$, and an Ekman layer of thickness $\delta_E \sim Ek^{1/2}$, the topography is larger than the Ekman layer by a factor of $\ep/\delta_E \sim 10^4$. 
We therefore choose topographic amplitudes that maintain this ordering of length scales in the system, i.e.~$\ep/\de \gg 1$, such that the topographic bump does not reside within the Ekman layer.}
%\textcolor{\revtxt}{However,}
\textcolor{\revtxt}{The majority of the simulations were performed with $\epsilon  = 0.2, 0.1$ in which this ordering is well satisfied, and more limited sweeps are carried out for $\epsilon  = [0.005,0.01,0.02,0.03,0.05,0.1,0.2,0.3].$ }

\subsection{\textcolor{\revtxt}{Asymptotic Scalings for Topographic Torques
}}
\label{s:tt}

\textcolor{\revtxt}{Insight into the dependence of the topographic torque on both the input parameter, $Ek$, and the output parameter, $Re$, can be gained by appealing to the asymptotic theory of rotating convection. Our goal is to provide only a brief summary that allows for understanding the results of the present study -- further details are provided in previous work \citep{mC13,mS06,jN24}. For a fixed value of the Ekman number, the flow is characterized by a geostrophic force balance between the Coriolis and pressure gradient forces provided that $\Rat$ is not too large \citep[e.g.][]{sM21}.
%\be
% \frac{2}{Ek}\mathbf{\hat{z} }\times \mathbf{u} \approx -\nabla p .
%\ee
The convective dynamics are then the result of perturbations, or small corrections, to this balance. We denote the relevant small parameter by $\delta \equiv Ek^{1/3}$, which characterizes the ratio of gradients in the axial (measured along the rotation axis) and horizontal (measured perpendicular to the rotation axis) directions.
Thus, the largest components of the gradient operator scale as $\nabla = O\lb \delta^{-1} \rb$ and the velocity field scales as $\mathbf{u} = O\lb \delta^{-1} \rb$. We define rescaled quantities as $\widetilde{\nabla} \equiv \delta \, \nabla = O(1)$  and $\widetilde{\mathbf{u}} \equiv \delta \, \mathbf{u} = O(1)$. A self-consistent set of perturbation equations can then be derived that allows for the possible importance of all terms (buoyancy, inertia, advection, and the viscous force) in the prognostic evolution equation\citep[][]{mC13,jN24}. The relative sizes of these perturbative terms is then determined by the strength of the buoyancy forcing, or $\Rat$. The requirement of leading order geostrophy implies that the pressure scales as
\be
p = O(\delta^{-3}) .
\ee
However, this scaling provides the asymptotic size of only the leading order, geostrophic component of the pressure. As alluded to in previous work\citep{aA77b,pR88}, the topographic torque is identically zero when the flow is completely geostrophic because it is constrained to follow lines of constant axial height, known as geostrophic contours \citep{hG68}. As we show here, it is then necessary to explicitly account for higher order corrections to the pressure. More specifically, the pressure is expanded in a perturbation series according to 
\be
%\ub = \delta^{-1} \ls \ub_0 + \delta \ub_1 + O\lb \delta^2 \rb \rs, \qquad  
p = \delta^{-3} \ls \widetilde{p}_0 + \delta \, \widetilde{p}_1 + O\lb \delta^2 \rb \rs,
\label{E:press}
\ee
This expansion, along with an analogous expansion for the velocity field, can then be inserted into the momentum equation and the leading order geostrophic balance arises as
\be
2\mathbf{\hat{z} }\times \widetilde{\mathbf{u}}_0 \approx -\widetilde{\nabla} \widetilde{p}_0 ,
\label{E:geo}
\ee
where $p_0$ is now explicitly identified as the geostrophic component of the pressure.
%We assume the hydrostatic component of the pressure has been removed via an appropriately defined averaging operator though the details are not important for the analysis presented here.
%At the next order all other forces enter such that
%\be
%\partial_t \ub_0+ \ub_0 \cdot \nabla \ub_0 + 2\mathbf{\hat{z} }\times \ub_1 = - \nabla
%\ee
For simplicity we focus on the approximate form of the torque, equation \eqref{E:torque}, which can be decomposed according to 
\be
\Gtz = \epsilon \oint h \, \partial_\phi p \, d a = 
\epsilon \delta^{-4}  I_0  +  
\ep \delta^{-3} I_1  +
O \lb \delta^{-2} \rb, 
\ee
where the two integrals are defined by
\be
I_0 = \oint h \,  \widetilde{\partial}_{\phi}  \widetilde{p}_0 \, da, \quad \text{and} \quad I_1 = \oint h \,  \widetilde{\partial}_{\phi}  \widetilde{p}_1 \, da.
\ee
The first integral is identically zero since geostrophic flow must follow geostrophic contours, implying linear dependence between $h$ and the geostrophic pressure.
%Using a cylindrical coordinate system $(s, \phi, z)$, we denote the axial height of the domain as $f(s,\phi)$, such that the requirement to follow the geostrophic contours is given by
%\be
%\ub_0 \cdot \nabla f = 0.
%\label{E:kin}
%\ee
%Provided that closed geostrophic contours exist in the system, this constraint is independent of $f(s,\phi)$. The geostrophic velocity, $\ub_0$, can be expressed in terms of the pressure gradient via \eqref{E:geo}, so that $\ub_0 = \lb - s^{-1} \partial_{\phi} p_0, \partial_s p_0, 0 \rb$. 
%The kinematic constraint \eqref{E:kin} then becomes
%\be
%\partial_s p_0 \, \partial_{\phi} f -  \partial_{\phi} p_0 \, \partial_s f = 0,
%\ee
%which states that the Jacobian, $J[p_0, f] = 0$. Thus, the vectors $\nabla p_0$ and $\nabla f$ are linearly dependent such that one vector can be written as a linear combination of the other. For instance, letting $\partial_{\phi} p_0 = c \, \partial_{\phi} h$, where $c$ is an arbitrary constant, allows the integral $I_0$ to be written as
%\be
%I_0 = c \oint \partial_\phi \lb  \frac{1}{2} h^2 \rb \, da = 0,
%\ee
%since $\phi$ is a periodic coordinate and $h$ is a single valued function.
Therefore, the asymptotic analysis tells us that the torque integral is of order
\be
\Gtz = \ep \delta^{-3} I_1 = O\lb \ep Ek^{-1} \rb.
\label{e:gtz_ek_sc}
\ee
This suggests that the reduced torque, $\Gstd Ek$, is an order one quantity in the limit $Ek \rightarrow 0$.}

\textcolor{\revtxt}{If we now consider the effect of flow speed, a second scaling involving the Reynolds number, $Re$, can be obtained. Since the primary contribution to the torque arises from perturbations to the geostrophic pressure, we suppose that $\widetilde{\nabla} \widetilde{p}_1$ must be comparable to other terms appearing in the prognostic equation governing the QG dynamics. For sufficiently large values of $\Rat$ we assume that nonlinearities become as large as pressure gradients. In the QG theory, $\widetilde{\ub}_{0}=O(Re \delta),$ so that
\be
\widetilde{\nabla} \widetilde{p}_1 \sim \widetilde{\ub}_0 \cdot \widetilde{\nabla} \widetilde{\ub}_0 = O\lb \delta^2 Re^2 \rb, \qquad \Rightarrow \qquad \widetilde{\partial}_{\phi}  \widetilde{p}_1 = O\lb Re^2 Ek^{2/3} \rb .
\ee
Using the above scaling in the integral $I_1$ then suggests that the torque scales as
\be
\Gtz = O\lb \ep Re^2 Ek^{-1/3} \rb .
\label{e:gtz_re_sc}
\ee
Equations \ref{e:gtz_ek_sc} and \ref{e:gtz_re_sc} can be reconciled by noting that $Re=O(Ek^{-1/3}) $ in the asymptotic theory.
}
%\textcolor{\revtxt}{
%	A theory for the empirical scaling $\Gstd\sim \epsilon Re^{2}Ek^{-1/3},$ is provided below.
%	%\textcolor{blue}{Previous work \cite{rH69,pR12} has suggested that $\Gamma_{z}$} . 
%The pressure is split into the hydrostatic component $p_{h}$ and a fluctuating component $p^{*},$ so that $p = p_{h} + p^{*}.$ $p^{*}$ is further expanded in an asymptotic series,
%\begin{equation}
%	p^{\prime} =  p_{0}^{*} + \delta p_{1}^{*} + ...,
%	\label{e:p_asy}
%\end{equation}
%where $\delta$ is the small parameter $Ek^{1/3}.$ The hydrostatic pressure is radial so that $\partial_{\phi}p_{s}=0$. The first term on the RHS of equation \ref{e:p_asy} is the geostrophic pressure; its gradient is given by
%\begin{equation}
%\frac{2}{Ek}\mathbf{\hat{z}} \times \mathbf{u} = -\nabla p_{0}^{*}.
%	\label{e:geo_p}
%\end{equation}
%}
%\textcolor{blue}{yuck...
%It can be shown that the geostrophic pressure does not contribute to the torque calculation because the isobars follow the geostrophic contours which are a function only of geometry and therefore steady in time.
%}
%\textcolor{\revtxt}{
%Therefore, the leading order contribution to the torque is $\delta^{2} p_{1}^{*}.$ Dynamical balance at $O\lb \delta\rb $ requires
%\[\mathbf u \cdot \nabla \mathbf{u} \sim -\delta \nabla p_{1}^{*}.\]
%In our non-dimensionalization, $\mathbf{u}\sim Re$ and the gradient operator in the horizontal directions scales as $Ek^{-1/3}.$ Therefore,
%\begin{equation}
%	\delta\;\partial_{\phi} p_{1}^{*} \sim Re^{2}Ek^{-1/3}\implies \Gamma_{z}\sim \epsilon  Re^{2}Ek^{-1/3}.
%	\label{e:scaling}
%\end{equation}
%}

\subsection{Relating $\widetilde{t}$ to length of day}

The rescaled time $\tt$ can be related to the length of day as follows. Denoting dimensional time (in seconds) as $\tau$ we have
\be 
\tt= Ek^{-2/3}\lb\frac{\nu }{d^2}\rb \tau = Ek^{1/3}\lb\frac{\nu}{d^2}\rb \lb\frac{\Omega d^2}{\nu}\rb\tau = Ek^{1/3}\Omega \tau.
\label{e:tt}
\ee
We note that $\tt$ should be considered as a small scale diffusion time. 
From the linear stability analysis $\ell/d = Ek^{1/3}$ where $\ell$ is the length scale associated with the convection at onset. 
Even in the turbulent regime, the flow structures, such as the thickness of the Taylor columns, may remain $O\lb Ek^{1/3}\rb $ \citep{tO23,jN24},
\textcolor{\revtxt}{
and convection experiments\citep{jA24} suggest that we are in a regime where the proposed $Ek^{1/3}$ and $Ro^{1/2}$ scalings\citep{cG19,jA20} may be indistinguishable}.
Therefore,
\be
\tt = \lb\frac{\nu}{\ell^2}\rb \tau.
\ee
Plugging in $Ek=10^{-15}$ and $\Omega = 7.27 \times 10^{-5}$ s$^{-1}$ into equation (\ref{e:tt}) gives 
\textcolor{\revttxt}{
\be
\tt = (7.27 \times 10^{-10} \text{s}^{-1})\cdot \tau .
\ee
}
A decadal timescale corresponds to \textcolor{\revtxt}{$\Delta\tau \approx 3 \times 10^{8}$} s such that
\textcolor{\revtxt}{
\be
\Delta\tt \approx 0.2.
\ee
}
%\textcolor{\revtxt}{The possibility of stably stratified layers regionalized near the LLSVP's has been proposed by previous work, for instance see Mound et al.\citep{jM19}. The existence of such regions would likely have a direct impact on the scaling of topographic torques. It is also of interest to consider the coupled effects of heat flux heterogeneity and topography.}
%	%but it is of interest to understand how the change in flow morphology, with smaller $Pr$ yielding larger zonal flows\citep{}, contributes to the torque. 
%	However, if the torques continue to be dominated by the convective fluctuations, then the effect of $Pr$ is expected only through the flow speeds ($Re$) which is directly considered in the present work.
%%We therefore expect our scaling to hold at geophysical values of $Pr$.
%}
%Thus, based on our state-of-the-art numerical simulations of convection in a rotating spherical shell, we conclude that kilometer-sized CMB topography can explain \textcolor{\revtxt}{the large amplitude} LOD variations.

\section*{Data availability}
The datasets presented in the current study are available at\\ 
\texttt{https://github.com/tgoliver/TGO\_Topographic\_Coupling\_Data/}\\
%Other data generated during this study are available from the corresponding author at reasonable request.
The data required to generate the visualizations are not available because they are too large to store on local machines.
\\
\\
\section*{References}
\bibliographystyle{naturemag}
\bibliography{References,journal_abbreviations}%jfm-bib,References}
\newpage

%\bibliographystylesupp{naturemag}
%\bibliographysupp{References,journal_abbreviations}%jfm-bib,References}
\section*{Acknowledgements}

\textcolor{\revtxt}{We would like to thank three anonymous reviewers for their helpful comments.} This work was funded by the National Science Foundation through CSEDI grant EAR-2201595 and grant EAR-1945270. The Stampede3 and Anvil supercomputers at the Texas Advanced Computing Center (TACC) and Purdue University, respectively, were made available through allocation PHY180013 from the Advanced Cyberinfrastructure Coordination Ecosystem: Services \& Support (ACCESS) program, which is supported by National Science Foundation grants \#2138259, \#2138286, \#2138307, \#2137603, and \#2138296 \citep{tB23}. Access to the Frontera supercomputer \citep{frontera} at TACC was made available through grant EAR24006. This work also utilized the Alpine high performance computing resource at the University of Colorado Boulder. Alpine is jointly funded by the University of Colorado Boulder, the University of Colorado Anschutz, and Colorado State University.
\\
\\
\section*{Author contributions}
T.G. Oliver prepared and performed the simulations. T.G. Oliver and M.A. Calkins prepared the text and figures. J.A. Tarduno and E.G. Blackman provided insight and assisted with text preparation.
\\
\\
\\
\\
\section*{Competing interests}
The authors declare no competing interests.
\section*{Additional Information}
Supplemental Figures are available for this paper.
\subsection*{Correspondence}
Correspondence and requests for materials should be addressed to Tobias G. Oliver.\\ 
\newpage
\section*{Supplementary Information}
\newpage
\includepdf[pages=1]{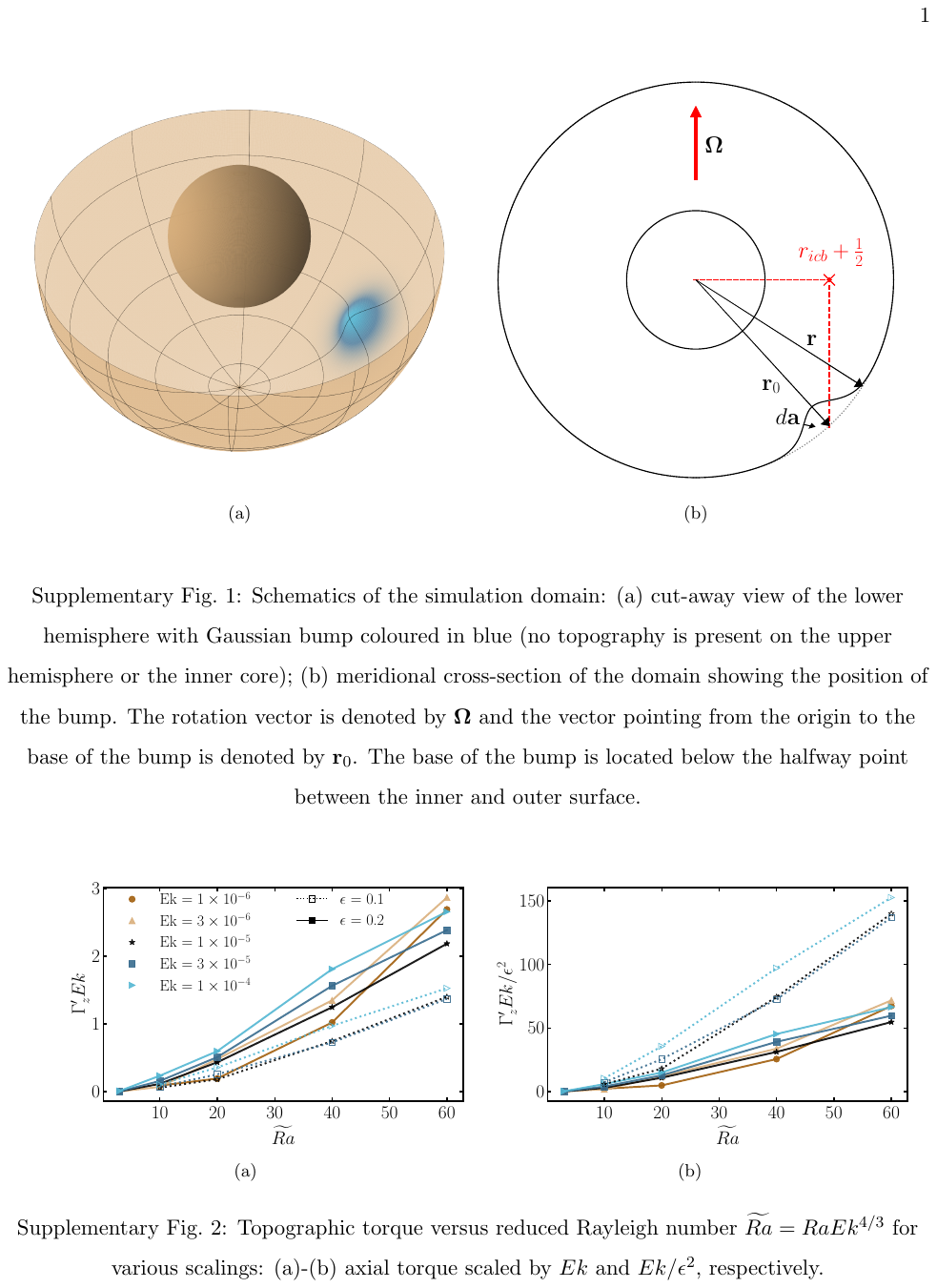}
\newpage
\includepdf[pages=2]{supp_ncee.pdf}

\end{document}